\documentclass[useAMS,usenatbib,referee]{biom}
\usepackage{graphicx,verbatim,array,multicol,fontenc,subfigure,pifont}
\usepackage{amsmath, amssymb, epsfig, color,mathrsfs,latexsym,fancyhdr,mathrsfs,pdflscape}
\usepackage{url}
\usepackage{amsbsy}
\usepackage{mathbbol}
\DeclareSymbolFontAlphabet{\amsmathbb}{AMSb}%

\def\bzero{{\bf 0}}

\def\ba{{\mbox{\boldmath$a$}}}
\def\bb{{\bf b}}

\def\be{{\bf e}}

\def\bq{{\bf q}}
\def\br{{\bf r}}

\def\bv{{\bf v}}
\def\bw{{\bf w}}

\def\bS{{\bf S}}

\def\bU{{\bf U}}

\def\bZ{{\bf Z}}

\def\thick#1{\hbox{\rlap{$#1$}\kern0.25pt\rlap{$#1$}\kern0.25pt$#1$}}

\def\bbeta{\boldsymbol{\beta}}

\def\bepsilon{\boldsymbol{\epsilon}}

\def\bDelta{\boldsymbol{\Delta}}

\def\smbalpha{\boldsymbol{{\scriptstyle{\alpha}}}}

\def\Ghat{{\widehat G}}

\def\bShat{{\widehat \bS}}

\def\betahat{{\widehat\beta}}

\def\bbetahat{{\widehat\bbeta}}

\def\smbalpha{\widehat{\smbalpha}}

\def\hbar{\bar{ h}}

\def\Hsc{{\cal H}}

\def\Lsc{{\cal L}}

\def\Qsc{{\cal Q}}

\def\Vsc{{\cal V}}

\def\Zsc{{\cal Z}}

\def\bBsc{\boldsymbol{\cal B}}

\def\bSsc{\boldsymbol{\cal S}}

\def\bBschat{\widehat{\boldsymbol{\cal B}}}

\def\real{{\mathbb R}}

\def\transpose{{\sf \scriptscriptstyle{T}}}

\def\half{\frac{1}{2}}

\def\nhalf{n^{\half}}

\def\nnhalf{n^{-\half}}

\def\E{\mbox{E}}

\def\var{\mbox{var}}

\def\sumin{\sum_{i=1}^n}

\def\trans{^{\transpose}}

\def\argmindum{\mathop{\mbox{argmin}}}
\def\argmin#1{\argmindum_{#1}}
\def\argmaxdum{\mathop{\mbox{argmax}}}
\def\argmax#1{\argmaxdum_{#1}}

\def\cov{\mbox{cov}}
\def\diag{\mbox{diag}}

\def\diag{\mbox{diag}}

\def\var{\mbox{var}}

\def\Cov{\mbox{Cov}}

\def\Var{\mbox{Var}}
\def\mybox#1{\vskip1mm \begin{center}
        \hspace{.0\textwidth}\vbox{\hrule\hbox{\vrule\kern6pt
\parbox{.9\textwidth}{\kern6pt#1\vskip6pt}\kern6pt\vrule}\hrule}
        \end{center} \vskip-5mm}
\def\lboxit#1{\vbox{\hrule\hbox{\vrule\kern6pt
      \vbox{\kern6pt#1\vskip6pt}\kern6pt\vrule}\hrule}}

\def\thickboxit#1{\vbox{{\hrule height 1mm}\hbox{{\vrule width 1mm}\kern6pt
          \vbox{\kern6pt#1\kern6pt}\kern6pt{\vrule width 1mm}}
               {\hrule height 1mm}}}

\def\Abb{\mathbb{A}}
\def\Bbb{\mathbb{B}}

\def\Fbb{\mathbb{F}}

\def\Ibb{\mathbb{I}}

\def\Kbb{\mathbb{K}}

\def\Obb{\mathbb{O}}
\def\Pbb{\mathbb{P}}

\def\Rbb{\mathbb{R}}

\def\Wbb{\mathbb{W}}

\def\bmath#1{\mbox{\boldmath$#1$}}
\def\fat#1{\hbox{\rlap{$#1$}\kern0.25pt\rlap{$#1$}\kern0.25pt$#1$}}

\def\Abb{\mathbb{A}}

\def\Wbb{\mathbb{W}}
\def\Wbbhat{\widehat{\mathbb{W}}}

\def\Tscr{\mathscr{T}}
\def\bTscr{\pmb{\Tscr}}
\def\Xscr{\mathscr{X}}
\def\Lscr{\mathscr{L}}
\def\Cscr{\mathscr{C}}
\def\Uscr{\mathscr{U}}
\def\Kbb{\mathbb{K}}
\def\bXscr{\pmb{\Xscr}}

\def\Nnhalf{N^{-\half}}

\def\sumiN{\sum_{i=1}^N}
\def\subSS{_{\scriptscriptstyle \sf SSL}}
\def\subdelta{_{\scriptscriptstyle \delta}}

\def\subS{_{\scriptscriptstyle S}}
\def\subdS{_{\scriptscriptstyle \delta,S}}

\def\subopt{_{\scriptscriptstyle\sf opt}}
\def\supb{^{\scriptscriptstyle\sf (b)}}
\def\sub{\scriptscriptstyle\sf}
\def\subdj{_{\scriptscriptstyle \delta,j}}
\def\Sigbb{\mathbb{\Sigma}}

\def\strans{^{*\top}}

\def\nhalf{n^{\half}}

\def\half{\frac{1}{2}}

\definecolor{darkred}{RGB}{150,50,50}
\definecolor{brown}{RGB}{250,100,100}
\definecolor{green}{RGB}{000,150,100}
\definecolor{purple}{RGB}{250,000,180}

\usepackage[dvipsnames]{xcolor}

\begin{document}

\author{Stephanie F. Chan$^{1\dag}$, Jue Hou$^{1\dag}$, Xuan Wang$^{1}$, and Tianxi Cai$^{1,2,*}$ \email{tcai@hsph.harvard.edu} \\
$^{1}$Department of Biostatistics, Harvard T.H. Chan School of Public Health, Boston, MA 02115\\
$^{2}$Department of Biomedical Informatics, Harvard Medical School, Boston, MA 02115\\
$^{\dag}$ Equal Contributors}
\title[Risk Prediction with Imperfect Outcome Information from EHRs]{Risk Prediction with Imperfect Survival Outcome Information from Electronic Health Records}

\begin{abstract}
Readily available proxies for time of disease onset such as time of the
first diagnostic code can lead to substantial risk prediction error if
performing analyses based on poor proxies. Due to the lack of detailed
documentation and labor intensiveness of manual annotation, it is often
only feasible to ascertain for a small subset the current status of the
disease by a follow up time rather than the exact time. In this paper, we
aim to develop risk prediction models for the onset time efficiently
leveraging both a small number of labels on current status and a large
number of unlabeled observations on imperfect proxies. Under a semiparametric transformation model for onset and a highly flexible
measurement error models for proxy onset time, we propose the semisupervised risk prediction method by combining information from proxies
and limited labels efficiently. From an initial estimator solely based on
the labelled subset, we perform a one-step correction with the full data
augmenting against a mean zero rank correlation score derived from the
proxies. We establish the consistency and asymptotic normality of the
proposed semi-supervised estimator and provide a resampling procedure
for interval estimation. Simulation studies demonstrate that the
proposed estimator performs well in finite sample. We illustrate the
proposed estimator by developing a genetic risk prediction model for
obesity using data from Partners Biobank Electronic Health Records
(EHR).


\end{abstract}

\maketitle

\section{Introduction}

Electronic health records (EHRs), containing detailed medical history of individuals in the health care system, hold immense potential for translational research  \citep{jensen2012mining}. In recent years, EHR data has been increasingly explored for
developing risk prediction models to assist in clinical decision making \citep[e.g.]{eapen2013validated,calvert2016using,jin2018predicting}.
The longitudinal  EHR data contain information on the occurrence time of
clinical events which can be used as outcomes for risk prediction modeling.
Rich clinical features, including lab measurements, medication prescriptions and co-morbidities,
can be extracted as risk factors. Such clinical information can be effectively extracted from either
codified data such as billing codes and procedure codes and from free text clinical notes via natural
language processing (NLP). At research institutions, EHR data have also been linked with biobanks
where genetic information can be included in addition to clinical features to further improve risk prediction.

Although longitudinal EHR data is of great value for risk prediction modeling, precise information
on clinical event time of interest, $T$, are not readily available.
Timing and number of diagnostic codes and mentions of the disease in the clinical notes
can serve as poxies of the true event time and status, but they are often not highly accurate.
For example,  at Partner's healthcare, having at least one diagnostic code of obesity only attained a
sensitivity of  73\% and positive predictive value (PPV) of 81\%; while having at least one NLP mention of
obesity attained a higher sensitivity of 92\% but a lower PPV of 68\%. It is even more challenging to
approximate event time. For lung cancer recurrence, time of chemotherapy and radiation
therapy initiations can only predict the true recurrence times up to 5.9 and 6.7 months on average
\citep{uno2018determining}. Directly using proxy event times $\bTscr$ to replace the true $T$ for risk modeling can
lead to substantial bias due to the measurement error in $\bTscr$. On the other hand, extracting
event time via manual annotation is too resource consuming  for large scale research. Furthermore, the
exact timing of the event may not be precisely documented in the record and hence it is  often only feasible
to annotate current status of the event $\delta = I(T \le C)$, i.e. whether the event has occurred by the end of
the follow-up $C$. In this paper, we aim to develop an efficient EHR-based risk prediction procedure under a
semi-supervised (SS) setting with data from a small set of gold standard labels on $\delta$ and a large
set of unlabeled data with mis-measured event times $\bTscr$.

With current status survival data, regression methods have proposed for commonly used survival models including the hazards (PH) and proportional odds (PO) models and the semi-parametric transformation model (STM) \citep{huang1996efficient,rossini1996semiparametric,huang1997sieve,van1998locally,CarrollEtal97,sun2005semiparametric}. Estimation procedures have also been proposed
for other models including the additive hazard model and accelerated failure time model \citep{lin1998additive, chen2010multiple, betensky2001computationally, tian2006accelerated}. 
Developing risk prediction models with mis-measured survival outcomes is a challenging problem and few methods currently exist, in part due to the additional complexity induced by censoring. Recently, \cite{oh2018considerations} introduced a bias correction procedure for hazard ratio estimates in the PH model.
Using a validation data where both the true and error prone survival times are both available, \cite{braunnonparametric} proposed a non-parametric bias correction procedure for models using $\bTscr$ as predictors.

No regression procedure currently exists for the (SS) setting with observations on both $\delta$ and $\bTscr$. In addition, even without censoring, existing methods incorporating mis-measured outcomes largely require restrictive parametric measurement error model assumptions. To fill this gap, we propose an efficient and robust SS estimation procedure for the STM under flexible measurement error models without distributional assumptions. Our SS estimation starts with an initial supervised estimator based on the current status data via solving a system of kernel smoothed estimating equations and then constructs an augmented estimator by optimally combining the initial estimator with information from $\bTscr$ derived from a rank estimation procedure. We analyze the limiting distribution of the proposed estimator and develop inference method accordingly. For the scenario of the unlabelled data is much larger than the labelled data, we discover an interesting ``space collapse'' phenomenon of the SS estimator,
which requires special arrangements for estimation and inference.

The rest of the paper is organized as follows. We detail our SS procedures in Section 2.
In Section 3, we present results from simulation studies to examine the finite sample behavior of the SS estimator and compare its efficiency to the initial supervised estimator. In Section 4, we apply our methods to develop an age-specific risk prediction model for obesity based on demographic and genetic information using EHR data from the Partners Biobank. Concluding remarks are giving in Section 5.
Technical details are in the appendices.

\section{Methods}

Suppose there are a total of $N$ subjects in the EHR cohort and a subset of $n$ subjects are randomly sampled into the labeled set to have their event status annotated via manual chart review. Let $T_i$, $C_i$ and $\bZ_i$ respectively denote the true event time, follow up time, and $p$ dimensional baseline covariates for the $i$th subject. The true event time $T_i$ is not observable but $\delta_i = I(T_i \le C_i)$ is observed for those in the labeled set. In addition, there are $K$ surrogate event times, $\bTscr = (\Tscr_1, ..., \Tscr_K)\trans$, that can be viewed as proxies of $T$. However, $\bTscr$ is subject to right censoring since patients are only followed in the EHR up to time $C$. Thus for $\bTscr$, we only observe $\bXscr = (\Xscr_1, ..., \Xscr_K)\trans$ and $\bDelta = (\Delta_1, ...,\Delta_K)\trans$, where $\Xscr_k = \Tscr_k \wedge C$ and $\Delta_k = I(\Tscr_k \le C)$.  The full underlying data of the EHR cohort consist of $ \{(T_i,C_i,\bZ_i\trans,\bTscr_i\trans)\trans , i= 1,\ldots,N \}$ while the observed data consist of the labeled data $\Lscr= \{(\delta_i,C_i,\bZ_i\trans,\bXscr_i\trans,\bDelta_i\trans)\trans , i= 1,\ldots,n \}$ and the unlabeled data $\Uscr= \{(C_i,\bZ_i\trans,\bXscr_i\trans,\bDelta_i\trans)\trans , i= n+1,\ldots,N \}$. Without loss of generality, we assume that $C$ has a continuous distribution with twice continuously differentiable density with finite support $[\Cscr_l, \Cscr_r]$.

We predict $T_i$ with $\bZ_i$ via the STM which includes PH and PO models as special cases:
\begin{equation}
P(T_i \leq t \mid \bZ_i) = g\{h_0(t) + \bbeta_0\trans\bZ_i\} \label{model-STM}
\end{equation}
where $g(\cdot)$ is a known smooth probability distribution function, $h_0(t)$ is an unspecified smooth increasing function, and $\bbeta_0$ is the unknown regression coefficient.
Under the STM \eqref{model-STM}, we have
\begin{equation}\label{eq:STM-moment}
  E(\delta_i \mid C_i, \bZ_i)
  = P(T_i \le C_i | C_i, \bZ_i)
  = g(h_0(C_i)+\bbeta_0\trans\bZ_i)
\end{equation}
from which we later derive our estimating equations for the initial estimator.

For each of the mis-measured survival outcome $\Tscr_{k}$, we assume that
\begin{equation}
\Hsc(\Tscr_{ki}) = \Hsc(T_i) + \epsilon_{ki}, \quad \mbox{for } k = 1, ...,K, \label{model-error}
\end{equation}
where and $\Hsc(\cdot)$ is an unknown smooth transformation function and $\epsilon_{ki}$ is independent of $(T_i, \bZ_i\trans, C_i)\trans$ with a completely unspecified distribution.
We also leave the within-subject correlation structure among $\bepsilon = (\epsilon_{1}, ..., \epsilon_{K})\trans$ unspecified. Leaving both $\Hsc(\cdot)$ and the distribution of $\epsilon_k$ unspecified allow a wide range of measurement error models, including both additive and multiplicative measurement errors.
With the finite observation window ending at $\Cscr_r$,
the truncated mis-measured survival outcome $\Tscr_{ki}^* = \Tscr_{ki} \wedge \Cscr_r$ is of greater practical interest,
as no event beyond $\Cscr_r$ is observable.
We show in \ref{app:sim} that $\Tscr_{ki}^*$ follows a generalized regression model as defined in
\citet{han1987non}. It follows from \citet{han1987non} and \citet{sherman1993limiting} that the direction of the true parameter
$\bbeta_0$, $\bBsc_0 = \bbeta_0/\|\bbeta_0\|_2$, is identified by the maximizer of the population rank correlation:
$$
\bBsc_0 = \argmax{\bbeta \in \Omega_p} \Qsc_k(\bbeta), \; \quad \mbox{where}\quad
\Qsc_k(\bbeta) = P\left(\bbeta\trans\bZ_i \ge \bbeta\trans\bZ_j \mid \Tscr_{ki}^* < \Tscr_{kj}^*\right)
$$
$\Omega_p = \{\bbeta \in \Rbb^p: \|\bbeta\|_2 = 1\}$ and  $\|\cdot\|_2$ denotes the $L_2$ norm.
From the first order KKT condition, 
\begin{equation}\label{eq:error-moment}
\bSsc_k(\bBsc_0) = \bzero , \quad \mbox{where}\quad
\bSsc_k(\bbeta) = E(\bZ_i - \bZ_j \mid \bbeta\trans\bZ_i = \bbeta\trans\bZ_j, \Tscr_{ki}^* < \Tscr_{kj}^*).
\end{equation}
Based on \eqref{eq:error-moment}, we derive our score for the full cohort,
through which we update the initial estimator with the information
from the mis-measured survival outcomes.

\subsection{Estimation}

To estimate $\bbeta_0$, we first obtain an initial estimator using the labeled data $\Lsc$ according to  the moment equation \eqref{eq:STM-moment}.
Specifically, we adopt the kernel smoothed estimation procedure and estimate $\{h_0(\cdot), \bbeta_0\}$ by solving
\begin{align}
 \sum_{i=1}^n  \bZ_i&[\delta_i - g\{h(C_i) + \bbeta\trans\bZ_i\}] = 0 \notag \\
 \sum_{i=1}^n  \Kbb_h(C_i - t) &[\delta_i - g\{h(t) + \bbeta\trans\bZ_i\}] = 0,
 \label{ee:init}
\end{align}
where $\Kbb_h(t) = h^{-1}\Kbb(t/h)$, $\Kbb(\cdot)$ is a smooth symmetric probability density function, and $h = n^{-\nu}$ with $\nu \in (1/4, 1/2)$. Let $\bbetahat\subdelta$ denote the resulting estimate for $\bbeta_0$.
From Remark 2 and Theorem 3 of \cite{CarrollEtal97}, we have the consistency and asymptotic normality of $\bbetahat\subdelta$.
We also show in  \ref{app:sim} that
\begin{align}\label{eq:init_asymp}
\sqrt{n}\{  \bbetahat\subdelta - \bbeta_0\} = & n^{-1/2} \sum_{i=1}^n  \bU_i + o_p(1)
\leadsto N\left(\mathbf{0}, \Bbb^{-1} \Sigbb\subdelta(\Bbb^{-1})\trans\right),
\end{align}
where $\bU_i = \Bbb^{-1}\{\bZ_i-\bb(C_j)\}\{\delta_i - g(\Zsc_i)\}$, $\Zsc_i = h_0(C_i) + \bbeta_0\trans\bZ_i$,
$\Bbb = \E\left[ \bZ_i\left\{\bZ_i - \bb(C_i)\right\}\trans g'(\Zsc_i)\right]$, $\bb(t) =  \ba_1(t)/a_0(t)$,
$a_k(t) = f_c(t) \E[\bZ_i^{\otimes k}g'\{h_0(t) + \bbeta_0\trans\bZ_i\}]$ with $\bZ^{\otimes 0} = 1$, $\bZ^{\otimes 1}=\bZ$
and $\bZ^{\otimes 2} = \bZ\bZ\trans$, $f_c(t)$ is the density function of  $C$, and
\begin{gather}
\Sigbb\subdelta = \E\left[\{\bZ_i-\bb(C_i)\}^{\otimes 2}  g(\Zsc_i) \{1- g(\Zsc_i)\}\right].  \label{def:BSigd}
\end{gather}
Since $\bbeta_0$ is often sparse, we can consistently recover the indices set of nonzero positions in $\bbeta_0$, $\mathrm{supp}(\bbeta_0) = \{j: \bbeta_{0,j} \neq 0\}$, by thresholding the initial estimator $\bbetahat\subdelta$,
 $$
 \widehat{\mathrm{supp}}(\bbeta_0) = \{j: |\bbeta_{0,j}| > \lambda\subdelta\}, \;
 \sqrt{n}\lambda\subdelta \to +\infty,
 $$
 which will be used in the construction of our final estimator.

To improve the efficiency of $\bbetahat\subdelta$ leveraging the mis-measured survival outcomes of the full cohort, we augment $\bbetahat\subdelta$ with an estimate for $\bSsc_k(\bBsc_0)$.
To this end, we estimate the rank correlation
score function (\ref{eq:error-moment}) via kernel smoothing as:
\begin{equation}\label{def:Sk}
  \bShat_k(\bBsc) =\frac{\sum_{i=1}^N\sum_{j = 1}^N(\bZ_i - \bZ_j)\Kbb_{h'}(\bBsc\trans\bZ_i - \bBsc\trans\bZ_j)\frac{I(\Xscr_{ki} < \Xscr_{kj})\Delta_{ki}}{\widehat{G}(\Xscr_{ki})^2}}{\sum_{i=1}^N\sum_{j = 1}^N\frac{I(\Xscr_{ki} < \Xscr_{kj})\Delta_{ki}}{\widehat{G}(\Xscr_{ki})^2}},
  \quad \mbox{for $k = 1, ..., K$},
\end{equation}
by noting that $$
 \E\left\{\frac{I(\Xscr_{ki} < \Xscr_{kj})\Delta_{ki}}{G(\Xscr_{ki})^2}\mid \Tscr_{ki}, \Tscr_{kj}\right\}
 = I(\Tscr_{ki}^* < \Tscr_{kj}^*).
 $$
 following \cite{CaiCheng08},
where $h' = N^{-\nu'}$ for $\nu' \in (1/4, 1/2)$ and $\Ghat(t)$ is the empirical estimate of $G(t) =P(C \ge t)$.
 We stack the scores evaluated at $\bBschat\subdelta$ across all $k = 1,\dots, K$ as $\bShat(\bBschat\subdelta)$, where
$\bShat(\bBsc) = [\bShat_1(\bBsc)\trans, ..., \bShat_K(\bBsc)\trans]\trans.$
Noting the connection between $\bShat_k(\bbeta)$ and the rank correlation
\begin{equation}\label{def:Qk}
   \widehat{Q}_k(\bBsc) =  N^{-2} \sum_{i=1}^N\sum_{j = 1}^N I(\bBsc\trans\bZ_i - \bBsc\trans\bZ_j > 0)\frac{I(\Xscr_{ki} < \Xscr_{kj})\Delta_{ki}}{\widehat{G}(\Xscr_{ki})^2}
\end{equation}
we derive in  \ref{app:Sk} the asymptotic distribution of $\widehat{\bS}(\bBschat\subdelta)$. Specifically, we show that
\begin{align}
   \nhalf \widehat{\bS}(\bBschat\subdelta)
 = & \nnhalf\sum_{i=1}^{n}2\Abb \|\bbeta_0\|_2^{-1} \bU_i  +\nhalf N^{-1}\sum_{i=1}^{N} 2\bU_{qi}
 + o_p\left(n^{-1/2} + N^{-1/2}\right),  \label{eq:Sk-bhat}
\end{align}
which converges in distribution to a zero-mean multivariate normal,
where $\Abb = \left(\Abb_1\trans,\dots, \Abb_K\trans\right)\trans$, $\Abb_k = \nabla^2 \E\{q_{k,i}(\bBsc_0)\}$,
$\bU_{qi} = \dot{\bq}_i(\bBsc_0)+ \dot{\bq}^c(C_i,\bBsc_0)$,
$\dot{\bq}_i(\bBsc) = \left(\dot{\bq}_{1,i}(\bBsc), \dots, \dot{\bq}_{K,i}(\bBsc)\right)\trans$,
$\dot{\bq}^c(c,\bBsc) = \left(\dot{\bq}^c_1(c,\bBsc), \dots, \dot{\bq}^c_K(c,\bBsc)\right)\trans$,
\begin{gather}
\begin{aligned}
\dot{\bq}_{k,i}(\bBsc) = \frac{1}{2}\frac{\partial}{\partial \bBsc}&
\E\left\{I(\bBsc\trans\bZ_i > \bBsc\trans\bZ_j)\frac{I(\Tscr_{ki} < \Tscr_{kj})\Delta_{ki}}{G(\Tscr_{ki})} \right. \\
& \left.\quad + I(\bBsc\trans\bZ_j > \bBsc\trans\bZ_i )\frac{I(\Tscr_{kj} < \Xscr_{ki})}{G(\Tscr_{kj})} \mid \bZ_i, \Xscr_{ki}, \Delta_{ki} \right\},
\end{aligned}\notag \\
  \begin{aligned}
   \dot{\bq}^c_k(c, \bBsc) =  \frac{1}{2}\frac{\partial}{\partial \bBsc}\E&  \left[I(\bBsc\trans\bZ_i> \bBsc\trans\bZ_j )I(\Tscr_{ki} < \Tscr_{kj}) \left\{\frac{I(c \ge \Tscr_{ki})}{G(\Tscr_{ki})} - 1\right\}\right. \notag \\
& \left.\quad + I(\bBsc\trans\bZ_j >\bBsc\trans\bZ_i)I(\Tscr_{kj} < \Tscr_{ki})\left\{\frac{I(c \ge \Tscr_{kj})}{G(\Tscr_{kj})} - 1\right\}\right],
   \end{aligned}   \label{def:qki_Ak_qkc}
\end{gather}

Our final estimator will construct optimal combination of $\bbeta\subdelta$ with $\bShat(\bbetahat\subdelta)$ to minimize.
The optimal combination strategy depends on the proportion of labelled data $n/N \to \rho$,
in part due to the co-linearity of $\nhalf\bShat(\bbetahat\subdelta)$, whose asymptotic covariance matrix is of rank at most $K(p-1)$.
We will next discuss the scenario of $n \asymp N$ and $n \ll N$ separately.

\subsubsection{Optimal Combination when $n \asymp N$}

In this scenario, the asymptotic covariance matrix of $\nhalf\bShat(\bbetahat\subdelta)$ is of rank $K(p-1)$.
To resolve the co-linearity in $\nhalf\bShat(\bbetahat\subdelta)$,
we drop the $j$-th element in $\widehat{\bS}_k(\bBschat\subdelta)$ for each $k=1,\dots, K$ and $j \in \mathrm{supp}(\bbeta_0)$. This can be done based on prior knowledge that a specific feature is predictive of the outcome.
We define the operation through the
following $K(p-1) \times Kp$ matrix
\begin{equation}\label{def:PbbIbb}
  \Pbb_j = \left( \begin{array}{ccc}
                 \Ibb_{p, -j}& \dots & \Obb_{(p-1)\times p} \\
                 \vdots & \ddots & \vdots \\
                 \Obb_{(p-1)\times p} & \dots & \Ibb_{p, -j}
               \end{array} \right),
  \Ibb_{p, -j} = \left( \begin{array}{cc}
                 \Ibb_{j-1} & \Obb_{(j-1)\times(p-j)} \\
                 \mathbf{0}_{j-1}\trans  & \mathbf{0}_{p-j}\trans \\
                 \Obb_{(p-j)\times (j-1)}  & \Ibb_{p-j}
               \end{array} \right),
\end{equation}
where $\Ibb_{s}=\diag\{1,..,1\}_{s\times s}$, $\mathbf{0}_s=(0,...,0)_{s\times 1}\trans$ and $\Obb_{s\times r} = [0]_{s\times r}$.
The optimal combination of $\bbetahat\subdelta$ and $\Pbb_j\bShat(\bbetahat\subdelta)$
is given by the projection
$$
    \bbetahat\subdelta -  \cov\{\bbetahat\subdelta, \Pbb_j\bShat(\bBschat\subdelta)\}\var\{\Pbb_j\bShat(\bBschat\subdelta)\}^{-1}\Pbb_j\bShat(\bbetahat\subdelta)
$$
while the projection $\cov\{\bbetahat\subdelta, \Pbb_j\bShat(\bBschat\subdelta)\}\var\{\Pbb_j\bShat(\bBschat\subdelta)\}^{-1}$ approaches asymptotically
\begin{align}
  \Wbb\subopt(\Pbb_j)\trans
  = &\left\{\Bbb^{-1} \Sigbb\subdelta(\Bbb^{-1})\trans\Abb\trans\|\bbeta_0\|_2^{-1}
     +\rho \Sigbb\subdS\right\}\Pbb_j\trans  \notag \\
     &\quad\times
     \left[2\Pbb_j\left\{\Abb\Bbb^{-1} \Sigbb\subdelta(\Bbb^{-1})\trans\Abb\trans\|\bbeta_0\|_2^{-2}
  + \rho (\Sigbb\subS+\Abb\Sigbb\subdS+\Sigbb\subdS\trans\Abb\trans)\right\}
  \Pbb_j\trans \right]^{-1}.    \label{eq:Wopt_Nsmall}
\end{align}
where $\rho = \lim_{n\to \infty} n/N$.
It is natural to combine the $p$ projections with $\Pbb_j$, $j=1,\dots,p$,
$$
\bbetahat\subdelta - \overline{\Wbb\subopt}\trans\bShat(\bbetahat\subdelta), \;
\overline{\Wbb\subopt}\trans =  \frac{1}{|\mathrm{supp}(\bbeta_0)|}\sum_{j\in \mathrm{supp}(\bbeta_0)} \Wbb\subopt(\Pbb_j)\trans\Pbb_j.
$$
Suppose $\widehat{\Wbb}(\Pbb_j)$ is a consistent estimator for $\Wbb\subopt(\Pbb_j)$,
we obtain the final estimator by
\begin{equation}
    \bbetahat\subSS = \bbetahat\subdelta - \Wbbhat_{\sub comb}\trans\bShat(\bbetahat\subdelta), \;
\Wbbhat_{\sub comb}\trans =  \frac{1}{|\widehat{\mathrm{supp}}(\bbeta_0)|}\sum_{j\in \widehat{\mathrm{supp}}(\bbeta_0)} \widehat{\Wbb}(\Pbb_j)\trans\Pbb_j.
\end{equation}
We show in Appendix XX that
\begin{align}
\nhalf(\bbetahat\subSS-\bbeta_0)
 = & 2\nnhalf\sum_{i=1}^{n}\left(\Ibb_p -   \overline{\Wbb\subopt}\trans\Abb \|\bbeta_0\|_2^{-1}\right) \bU_i  -2\rho^{\half}\Nnhalf\sum_{i=1}^{N}  \overline{\Wbb\subopt}\trans\bU{qi} + o_p(1)\\
 \leadsto
  & N\left(\mathbf{0},4\Sigbb\subSS\right), \label{eq:SSL_norm}
\end{align}
where
\begin{gather}
  \begin{aligned}
\Sigbb\subSS = &  \left(\Ibb_p -  \overline{\Wbb\subopt}\trans\Abb \|\bbeta_0\|_2^{-1}\right)\Bbb^{-1} \Sigbb\subdelta(\Bbb^{-1})\trans\left(\Ibb_p +  \overline{\Wbb\subopt}\trans\Abb \|\bbeta_0\|_2^{-1}\right)\trans \\
&  + \rho \left\{ \overline{\Wbb\subopt}\trans\Sigbb\subS \overline{\Wbb\subopt}-\left(\Ibb_p - \overline{\Wbb\subopt}\trans\Abb \|\bbeta_0\|_2^{-1}\right)\Sigbb\subdS \overline{\Wbb\subopt}- \overline{\Wbb\subopt}\trans\Sigbb\subdS\trans\left(\Ibb_p - \overline{\Wbb\subopt}\trans\Abb \|\bbeta_0\|_2^{-1}\right)\trans\right\}.
  \end{aligned}
  \notag
\end{gather}

\subsubsection{Optimal Combination when $n \ll N$}
In this scenario, the asymptotic covariance matrix of $\nhalf\bShat(\bbetahat\subdelta)$ is rank $p-1$.
To resolve the co-linearity in $\nhalf\bShat(\bbetahat\subdelta)$,
we keep $\widehat{\bS}_k(\bBschat\subdelta)$ for one $k \in \{1,\dots, K\}$
and drop its $j$-element for some $j \in \mathrm{supp}(\bbeta_0)$.
We define the operation through the following $(p-1)\times K(p-1)$ matrix
\begin{equation}\label{def:Pbbkj}
  \Pbb_{k,j} = \left(\Obb_{(p-1)\times (pj-p)}, \Ibb_{p, -j}, \Obb_{(p-1)\times (p^2-pj)}\right),
\end{equation}
where $\Obb_{s\times r}$ and $\Ibb_{p, -j}$ are defined with \eqref{def:PbbIbb}.
Likewise, the optimal combination of $\bbetahat\subdelta$ and $\Pbb_j\bShat(\bbetahat\subdelta)$
is given by the projection
$$
    \bbetahat\subdelta -  \cov\{\bbetahat\subdelta, \Pbb_{k,j}\bShat(\bBschat\subdelta)\}\var\{\Pbb_{k,j}\bShat(\bBschat\subdelta)\}^{-1}\Pbb_{k,j}\bShat(\bbetahat\subdelta)
$$
while the projection $\cov\{\bbetahat\subdelta, \Pbb_{k,j}\bShat(\bBschat\subdelta)\}\var\{\Pbb_{k,j}\bShat(\bBschat\subdelta)\}^{-1}$ approaches asymptotically
\begin{equation}\label{eq:Wopt_Nlarge}
 \Wbb\subopt(\Pbb_{k,j})\trans
  =  \Bbb^{-1} \Sigbb\subdelta(\Bbb^{-1})\trans\Abb\trans\Pbb\trans
  \left\{2\Pbb\Abb\Bbb^{-1} \Sigbb\subdelta(\Bbb^{-1})\trans\Abb\trans\Pbb\trans\right\}^{-1}\Pbb\|\bbeta_0\|_2
\end{equation}
We combine the $Kp$ projections with $\Pbb_{k,j}$,$k=1,\dots,K$, $j=1,\dots,p$,
$$
\bbetahat\subdelta - \overline{\Wbb\subopt}\trans\bShat(\bbetahat\subdelta), \;
\overline{\Wbb\subopt}\trans =  \frac{1}{K|\mathrm{supp}(\bbeta_0)|}\sum_{k=1}^K\sum_{j\in \mathrm{supp}(\bbeta_0)} \Wbb\subopt(\Pbb_{k,j})\trans\Pbb_{k,j}.
$$
Suppose $\widehat{\Wbb}(\Pbb_{k,j})$ is an estimator for $\Wbb\subopt(\Pbb_{k,j})$,
we obtain the final estimator by
\begin{equation}
    \bbetahat\subSS = \bbetahat\subdelta - \Wbbhat_{\sub comb}\trans\bShat(\bbetahat\subdelta), \;
\Wbbhat_{\sub comb}\trans =  \frac{1}{K|\widehat{\mathrm{supp}}(\bbeta_0)|}\sum_{k=1}^K\sum_{j\in \widehat{\mathrm{supp}}(\bbeta_0)} \widehat{\Wbb}(\Pbb_{k,j})\trans\Pbb_{k,j}.
\end{equation}
We derive the asymptotic approximation of $\bbetahat\subSS$ as
\begin{equation}\label{eq:SSL-Nggn}
 \bbetahat\subSS-\bbeta_0 =  \frac{2}{n}\sum_{i=1}^{n}\frac{\bbeta_0\bbeta_0\trans \Bbb\trans \Sigbb\subdelta^{-1}\Bbb}{\bbeta_0\trans \Bbb\trans \Sigbb\subdelta^{-1}\Bbb\bbeta_0} \bU_i
  + o_p\left(n^{-1/2}\right).
\end{equation}
Thus, the asymptotic distribution of $\bbetahat\subSS$ is
\begin{equation}
 \sqrt{n}(\bbetahat\subSS-\bbeta_0)
 \leadsto N\left(\mathbf{0}, \frac{4\bbeta_0\bbeta_0\trans}{\bbeta_0\trans \Bbb\trans \Sigbb\subdelta^{-1}\Bbb\bbeta_0}\right).
\end{equation}

\subsection{Inference}
In both scenarios, the estimation of the optimal combination boils down to the estimation
of $\Wbb\subopt(\Pbb)$ with operation $\Pbb$.
We propose a resampling based procedure \citep{JinYingWei01}. Specifically, for $b = 1, ..., B$ with $B \gg n$, we generate a vector of $N$ independent and identically distributed random variables $\{\Vsc_i\supb, i = 1, ..., N\}$ with $E(\Vsc_i) = \var(\Vsc_i) = 1$ and obtain $\bbetahat\subdelta\supb$ as the solution to
\begin{align*}
 \sum_{i=1}^n  \Kbb_h(C_i - t) [\delta_i - g\{h_0(t) + \bbeta\trans\bZ_i\}] \Vsc_i\supb &= 0 , \\
 \sum_{i=1}^n  \bZ_i[\delta_i - g\{h_0(C_i) + \bbeta\trans\bZ_i\}] \Vsc_i\supb&= 0 .
\end{align*}
Then we obtain the perturbed counterpart of $\bShat_k(\bbeta)$ as
$$
\bShat_k\supb(\bBsc) = \frac{\sum_{1 \le i <  j \le N}\Vsc_i\supb\Vsc_j\supb(\bZ_i - \bZ_j)K_{h'}(\bBsc\trans\bZ_i - \bBsc\trans\bZ_j)\frac{I(\Xscr_{ki} < \Xscr_{kj})\Delta_{ki}}{\widehat{G}\supb(\Xscr_{ki})^2}}{\sum_{1 \le i <  j \le N}\Vsc_i\supb\Vsc_j\supb\frac{I(\Xscr_{ki} < \Xscr_{kj})\Delta_{ki}}{\widehat{G}\supb(\Xscr_{ki})^2}}, \quad
\mbox{for $k = 1, ..., K$},
$$
where $\Ghat\supb(t) = \{\sumiN I(C_i\ge t)\Vsc_i\supb\}/(\sumiN \Vsc_i\supb)$. Subsequently, we obtain $\bBschat\subdelta = \bbetahat\supb\subdelta/\|\bbetahat\supb\subdelta\|_2$ and $\bShat\supb(\bBschat\subdelta\supb)=[\bShat_1\supb(\bBschat\subdelta\supb)\trans, ..., \bShat_K\supb(\bBschat\subdelta\supb)\trans]$.
We consider the least square problem
\begin{equation}\label{eq:lm_what}
  \widehat{\bw}_j(\Pbb) = \argmin{\bw \in \real^{r}}\frac{1}{B}\sum_{b=1}^B \left\{\betahat\subdj\supb - \bw\trans \Pbb\bShat\supb(\bBschat\subdelta\supb)\right\}^2 ,
\end{equation}
where $r$ is the row dimension of $\Pbb$.
We assemble the estimated coefficients $\widehat{\bw}_1(\Pbb), \dots, \widehat{\bw}_{p}(\Pbb)$
to construct estimated projection matrix
\begin{equation}\label{eq:What-rate}
  \Wbbhat(\Pbb) =  \Pbb\trans (\widehat{\bw}_1, \dots, \widehat{\bw}_{p})
  = \Wbb\subopt(\Pbb) + O_p\left(n^{-1/2}\right).
\end{equation}
By \eqref{eq:What-rate} and
the consistency of support recovery $\widehat{\mathrm{supp}}(\bbeta_0)$, we have $ \Wbbhat_{\sub comb}=  \overline{\Wbb\subopt} + O_p\left(n^{-1/2}\right)$.

To quantify the estimation error,
it is natural to consider the perturbed SSL estimator
$$
\bbetahat\supb\subSS = \bbetahat\supb\subdelta - \Wbbhat_{\sub comb}\trans\bShat\supb(\bBschat\subdelta\supb), \; b = 1, ..., B.
$$
By \cite{JinYingWei01}, $\bShat\supb(\bBschat\subdelta\supb)$ has the same asymptotic distribution of
$\bShat(\bBschat\subdelta)$.
From \eqref{eq:Sk-bhat}, we have $\bShat\supb(\bBschat\subdelta\supb) = O_p(n^{-1/2})$.
Since $\bShat\supb(\bBschat\subdelta\supb)$ converges to zero,
the variability from estimating $\overline{\Wbb\subopt}$ by $\Wbbhat_{\sub comb}$
has negligible effect on the asymptotical distribution of $\bbetahat\supb\subSS$.
Hence, $\Wbbhat_{\sub comb}$ can be used directly in the perturbed SSL estimator
without an extra layer of perturbation.
Let $Q_\alpha(\bv)$ be the $\alpha$-quantile of
the perturbations $\{\bv\trans\bbetahat\supb\subSS: b = 1,\dots, B\}$.
Observing the skewness in the finite sample distribution of $\bbetahat\supb\subSS$,
 we recommend the following $(1-\alpha) \times 100 \%$ confidence interval
 for $\bv\trans \bbeta_0$ with re-centered empirical quantiles
\begin{equation}\label{def:CI-reg}
\left[Q_{\alpha/2}(\bv)+ \bv\trans\bbetahat\subSS-\frac{1}{B}\sum_{b=1}^{B}\bv\trans\bbetahat\supb\subSS,
Q_{1-\alpha/2}(\bv)+ \bv\trans\bbetahat\subSS-\frac{1}{B}\sum_{b=1}^{B}\bv\trans\bbetahat\supb\subSS\right].
\end{equation}

In the $N \gg n$ scenario, we discover an interesting ``space collapse'' phenomenon from the representation
\eqref{eq:SSL-Nggn}.
The SSL estimation error $\bbetahat\subSS-\bbeta_0$ concentrates in a one-dimensional subspace in $\Rbb^p$ spanned by
$\bbeta_0$ with large probability.
For any vector $\bv$ such that $\bv\trans\bbeta_0 = 0$, the SSL estimator $\bv\trans\bbetahat\subSS$
is super efficient,
$$
\bv\trans\bbetahat\subSS = o_p\left(n^{-1/2}\right).
$$
A typical example is the zero element in $\bbeta_0$ indexed by $j \notin \mathrm{supp}(\bbeta_0)$,
which can be represented with $j$-th natural basis $\be_j$,
$$
\be_j\trans\bbeta = \beta_{0,j} = 0, \;
\be_j\trans\bbetahat\subSS = \hat{\beta}_{\scriptscriptstyle SSL, j} = o_p\left(n^{-1/2}\right).
$$
Since $\sqrt{n}(\hat{\beta}_{\scriptscriptstyle SSL, j}-0)$ is no longer asymptotically regular,
the perturbation samples $\sqrt{n}\hat{\beta}_{\scriptscriptstyle SSL, j}\supb$ might not capture its asymptotic distribution.
To provide a valid inference for $\beta_{0,j}$, we perform the norm preserving soft-thresholding,
\begin{gather*}
\bbetahat_{\sub soft} = (\hat{\beta}_{\sub soft,1}, \dots, \hat{\beta}_{\sub soft,p})\trans, \;
  \hat{\beta}_{\sub soft,j} = \mathrm{sign}(\hat{\beta}_{\sub SSL,j})\max\{|\hat{\beta}_{\sub SSL,j}|-\lambda_{\sub soft}/|\hat{\beta}_{\sub SSL,j}|,0\},
  \\
\bbetahat_{\sub soft}\supb = (\hat{\beta}_{\sub soft,1}\supb, \dots, \hat{\beta}_{\sub soft,p}\supb)\trans, \;
\hat{\beta}_{\sub soft,j}\supb = \mathrm{sign}(\hat{\beta}_{\sub SSL,j}\supb)\max\{|\hat{\beta}_{\sub SSL,j}\supb|-\lambda_{\sub soft}/|\hat{\beta}_{\sub SSL,j}\supb|, 0\}, \\
\bbetahat_{\sub std} = \bbetahat_{\sub soft}\frac{\|\bbetahat\subSS\|_2}{\|\bbetahat_{\sub soft}\|_2}, \;
\bbetahat_{\sub std}\supb = \bbetahat_{\sub soft}\supb\frac{\|\bbetahat\subSS\supb\|_2}{\|\bbetahat_{\sub soft}\supb\|_2}
\end{gather*}
for $ n^{-1/2}\gg \lambda_{\sub soft} \gg n^{-1}$.
With the soft-thresholding, we also require that the re-centering of $\bbetahat_{\sub std}\supb$ preserves the sign,
$$
\hat{\beta}_{\sub center, j}\supb = \mathrm{sign}(\hat{\beta}_{\sub std, j}\supb)\max\left\{|\hat{\beta}_{\sub std, j}\supb|+ \mathrm{sign}(\hat{\beta}_{\sub std, j}\supb)\left(\hat{\beta}_{\sub std, j}-\frac{1}{B}\sum_{b=1}^{B}\hat{\beta}_{\sub std, j}\supb\right),0\right\}
$$
Let $Q_{std,j,\alpha}$ be the $\alpha$-quantile of
the perturbations $\{\hat{\beta}_{\sub center, j}\supb : b = 1,\dots, B\}$.
 We construct the $(1-\alpha) \times 100 \%$ confidence interval
 for $\beta_{0,j}$ with $\left[Q_{std,j,\alpha/2},
Q_{std,j,1-\alpha/2}\right]$.
Since we have $\hat{\beta}_{\sub SSL, j}$ is asymptotically regular and
$\hat{\beta}_{\sub center, j}\supb$ is asymptotically equivalent to $\hat{\beta}_{\sub SSL, j}\supb$
$$
\hat{\beta}_{\sub center, j}\supb = \hat{\beta}_{\sub SSL, j}\supb + O(\lambda_{\sub soft})
=  \hat{\beta}_{\sub SSL, j}\supb + o(n^{-1/2}),
$$
the confidence interval for $\bbeta_{0,j} \neq 0$
with the re-centered  norm preserving soft-thresholded perturbation
must achieve asymptotically the nominal coverage.
The confidence interval for $\bbeta_{0,j} = 0$,
however,
would have $100 \%$ coverage rate asymptotically as the  $\hat{\beta}_{\sub soft,j}$ and $\hat{\beta}_{\sub soft,j}\supb$, $b=1, \dots, B$ are all shrunk to zero with large probability.

We recommend select $\lambda_{\sub soft}$ through a cross-validation scheme targeting the aggregated rank correlation of
mis-measured survival times
\begin{equation*}
  \hat{\Qsc}(\bbeta) = \sum_{k=1}^K\hat{\Qsc}_k(\bbeta)
  = \sum_{k=1}^K \frac{\sum_{i=1}^N\sum_{j = 1}^N\frac{I(\bbeta\trans\bZ_i > \bbeta\trans\bZ_j)I(\Xscr_{ki} < \Xscr_{kj})\Delta_{ki}}{\widehat{G}(\Xscr_{ki})^2}}
  {\sum_{i=1}^N\sum_{j = 1}^N\frac{I(\Xscr_{ki} < \Xscr_{kj})\Delta_{ki}}{\widehat{G}(\Xscr_{ki})^2}}.
\end{equation*}
Exact evaluation of higher order terms in \eqref{eq:SSL-Nggn} is possible
but would multiply the computational burden.
Since testing $\beta_{0,j} = 0$ can be achieved solely in an unsupervised fashion with the mis-measured survival times based
on the rank correlation \citep{CaiCheng08}, we choose not to deviate from our SSL framework and expand in such direction.

\section{Simulations}

We conducted simulation studies to assess the performance of our estimator in finite sample settings. Throughout, we set the labelled sample size as $n=500$ and consider two sizes for total sample
$N = 1000$ and $N = 10000$.
The two total sample sizes characterize the $n \asymp N$ and $n \ll N$ scenarios,
respectively.
We first generated a $p=10$ dimensional $\bZ$ from a zero-mean multivariate normal with unit variance and correlation 0.2. Then we generated $T_i$ from
$$
3 \log(T_i/4) = - 0.7(Z_{i1}+Z_{i2}+Z_{i3}) + 0.5(Z_{i4}+Z_{i5}+Z_{i6}) - 0.3(Z_{i7}+Z_{i8}+Z_{i9}) + \varepsilon_i
$$
where $\varepsilon_i \sim N(0,1)$. The censoring $C_i$ was generated from from Uniform$(0,a)$ with $a$ chosen such that
$P(\delta_i = 1) = 0.5$. We consider $K=2$ surrogates generated from $\log \Tscr_{ki} = \log T_i + \epsilon_{ki}$, where
$\epsilon_k$ is generated from a normal mixture $\sim D_{ki} N(\mu_{k-}, \sigma_{k-}^2) + (1-D_{ki})N(\mu_{k+}, \sigma_{k+}^2)$ and $D_{ki} \sim \mbox{Bernoulli}(0.5)$. The normal mixture measurement error  distribution was chosen to allow for more heterogeneity in how their records appear in the EHR system.
We consider two scenarios for the error distribution: (A) low measurement error with
$(\mu_{1-},\sigma_{1-},\mu_{1+},\sigma_{1+}) = (0.2,0.3,-0.1,0.1)$ and
$(\mu_{2-},\sigma_{2-},\mu_{2+},\sigma_{2+}) = (0,0.2,0.3,0.1)$; (B) high measurement error with
$(\mu_{1-},\sigma_{1-},\mu_{1+},\sigma_{1+}) = (1,1.5,-0.5,0.5)$ and
$(\mu_{2-},\sigma_{2-},\mu_{2+},\sigma_{2+}) = (0,1,1.5,0.5)$.

The bandwidth $h$ for estimating $\bbeta\subdelta$ was chosen to be $\widehat{\tau} (\sumin \delta_i)^{-0.25}$, where $\widehat{\tau}$ is the empirical standard deviation of $C$. The bandwidth $h_k'$ for $\bShat_k$ is chosen to be $\widehat{\sigma}  (\sumiN \Delta_{ki})^{-0.3}$ where $\widehat{\sigma}$ is the empirical standard error of $\bBschat\subdelta\trans\mathbf{Z}$. For each scenario, we summarize results using 500 datasets. For the standard error estimates via resampling, we use $B=200$ replications.

In Table \ref{tab:re}, we show results for the bias, mean square error (MSE) of $\bbetahat\subSS$ and the relative efficiency (RE) of $\bbetahat\subSS$ compared to the supervised estimator $\bbetahat\subdelta$. Both estimators have negligible biases across all settings.
When $N \asymp n$,  the SS estimator $\bbetahat\subSS$ is more efficient than the supervised estimator $\bbeta\subdelta$ under the low measurement error model with efficiency gain ranging from 27\% to 62\% but the efficiency gain is minimal for the high measurement error setting. When $N \gg n$, $\bbetahat\subSS$ is substantially more efficient than $\bbeta\subdelta$ with efficiency gain ranging from 185\% to 9964\% for the low measurement error setting and from 133\% to 1559\% for the high measurement error setting. The efficiency gain is the highest for the zero coefficient as expected from our theoretical findings on the super efficiency of $\bbetahat\subSS$ on zero coefficients. These results also suggest that
the semi-supervised learning improves efficiency even with large measurement error in surrogates
when the unlabelled data is large.

\begin{table}[ht]
\begin{center}
\caption{Bias ($\times 100$) of $\hat{\beta}_\delta$ and $\hat{\beta}_{\rm SSL}$
 as well as relative effciency (RE) of $\hat{\beta}_{\rm std}$
 compared to $\hat{\beta}_\delta$ when the measurement errors in
$\mathcal{T}_1$ and $\mathcal{T}_2$ are (A) moderate and
(B) high.
} \label{tab:re}
\begin{tabular}{l|l|ll|ll}
\hline
\hline
\multicolumn{6}{c}{n=500, N = 500}\\
\hline
&& \multicolumn{2}{c|}{(A) Small}& \multicolumn{2}{c}{(C) High} \\
\cline{3-6}
& $\rm Bias_{\delta}$ & $\rm Bias_{\rm SSL}$ & RE & $\rm Bias_{\rm SSL}$ & RE \\
  \hline
$\beta_{1}$ & -1.82 & -2.27 & 1.38 & -1.91 & 0.95 \\
  $\beta_{2}$ & -3.96 & -3.70 & 1.27 & -1.78 & 1.01 \\
  $\beta_{3}$ & -2.30 & -2.43 & 1.27 & -4.21 & 1.09 \\
  $\beta_{4}$ &  1.92 &  3.05 & 1.39 &  3.63 & 1.07 \\
  $\beta_{5}$ &  2.83 &  3.92 & 1.58 &  3.87 & 1.02 \\
  $\beta_{6}$ &  1.68 &  3.24 & 1.40 &  2.69 & 1.04 \\
  $\beta_{7}$ & -0.91 & -0.81 & 1.58 & -0.31 & 1.16 \\
  $\beta_{8}$ & -0.63 & -0.99 & 1.58 & -1.23 & 1.15 \\
  $\beta_{9}$ & -1.90 & -1.02 & 1.62 & -0.47 & 1.22 \\
  $\beta_{10}$ &  0.14 &  1.38 & 1.57 &  1.55 & 1.11 \\
   \hline

\multicolumn{6}{c}{n=500, N = 10000}\\
\hline
&& \multicolumn{2}{c|}{(A) Small}& \multicolumn{2}{c}{(C) High} \\
\cline{3-6}
& $\rm Bias_{\delta}$ & $\rm Bias_{\rm SSL}$ & RE & $\rm Bias_{\rm SSL}$ & RE \\
  \hline
$\beta_{1}$ & -1.60 &  0.48 &   2.85 & -0.21 &  2.33 \\
  $\beta_{2}$ & -2.05 &  0.23 &   3.34 & -0.01 &  2.58 \\
  $\beta_{3}$ & -2.12 &  0.36 &   3.15 &  0.17 &  2.40 \\
  $\beta_{4}$ &  1.04 & -0.10 &   5.22 &  0.86 &  3.54 \\
  $\beta_{5}$ &  3.03 & -0.22 &   5.57 &  0.70 &  3.22 \\
  $\beta_{6}$ &  0.65 & -0.15 &   5.98 &  0.68 &  3.60 \\
  $\beta_{7}$ & -1.06 & -0.80 &   9.18 & -1.39 &  4.33 \\
  $\beta_{8}$ & -0.94 & -0.57 &   9.75 & -0.56 &  4.64 \\
  $\beta_{9}$ & -0.85 & -0.61 &  10.61 & -1.78 &  4.80 \\
  $\beta_{10}$ & -0.06 &  0.14 & 100.64 &  0.40 & 16.59 \\
   \hline

\hline
\end{tabular}
\end{center}
\end{table}

We also investigated the performance of our resampling procedure for variance and interval estimation. As shown in Table \ref{tab:cov}, for individual components of the regression coefficients, the average estimated standard errors are close to the corresponding empirical standard errors and the empirical coverage levels of the 95\% confidence intervals are close to the nominal level for nonzero coefficients.

\begin{table}[ht]
\begin{center}
\caption{Empirical SE (ESE) of norm-preserving soft-thresholded SSL estimator $\bbetahat_{\sub std}$, average of the estimated SEs (ASE) from norm-preserving soft-thresholded SSL perturbation $\bbetahat_{\sub soft, std}\supb$,
 empirical coverage levels (CovP) of the quantile based 95\% CIs
from re-centered norm-preserving soft-thresholded SSL perturbation $\bbetahat_{\sub center}\supb$.
} \label{tab:cov}
\begin{tabular}{l|lll|lll}
\hline
\hline
\multicolumn{7}{c}{n=500, N = 1000}\\
\hline
& \multicolumn{3}{c|}{(A) Small}&\multicolumn{3}{c}{(C) High} \\
\cline{2-7}
& ESE & ASE & CovP  & ESE & ASE & CovP  \\
  \hline
$\beta_{1}$ & 0.1275 & 0.1326 & 0.9420 & 0.1533 & 0.1441 & 0.9260 \\
  $\beta_{2}$ & 0.1331 & 0.1323 & 0.9380 & 0.1497 & 0.1443 & 0.9380 \\
  $\beta_{3}$ & 0.1329 & 0.1331 & 0.9480 & 0.1432 & 0.1439 & 0.9320 \\
  $\beta_{4}$ & 0.1223 & 0.1258 & 0.9340 & 0.1395 & 0.1380 & 0.9340 \\
  $\beta_{5}$ & 0.1176 & 0.1250 & 0.9360 & 0.1460 & 0.1378 & 0.9120 \\
  $\beta_{6}$ & 0.1282 & 0.1254 & 0.9360 & 0.1487 & 0.1390 & 0.9280 \\
  $\beta_{7}$ & 0.1169 & 0.1208 & 0.9600 & 0.1364 & 0.1334 & 0.9480 \\
  $\beta_{8}$ & 0.1182 & 0.1212 & 0.9420 & 0.1385 & 0.1338 & 0.9460 \\
  $\beta_{9}$ & 0.1139 & 0.1196 & 0.9540 & 0.1311 & 0.1336 & 0.9580 \\
  $\beta_{10}$ & 0.1142 & 0.1179 & 0.9480 & 0.1355 & 0.1317 & 0.9280 \\
   \hline

\multicolumn{7}{c}{n=500, N = 10000}\\
\hline
& \multicolumn{3}{c|}{(A) Small}&\multicolumn{3}{c}{(C) High} \\
\cline{2-7}
& ESE & ASE & CovP & ESE & ASE & CovP  \\
  \hline
$\beta_{1}$ & 0.0869 & 0.0854 & 0.9400 & 0.0962 & 0.0988 & 0.9440 \\
  $\beta_{2}$ & 0.0859 & 0.0850 & 0.9540 & 0.0978 & 0.0986 & 0.9400 \\
  $\beta_{3}$ & 0.0847 & 0.0850 & 0.9360 & 0.0971 & 0.0992 & 0.9480 \\
  $\beta_{4}$ & 0.0653 & 0.0678 & 0.9440 & 0.0788 & 0.0857 & 0.9640 \\
  $\beta_{5}$ & 0.0625 & 0.0662 & 0.9580 & 0.0819 & 0.0854 & 0.9620 \\
  $\beta_{6}$ & 0.0616 & 0.0682 & 0.9580 & 0.0791 & 0.0856 & 0.9580 \\
  $\beta_{7}$ & 0.0463 & 0.0519 & 0.9540 & 0.0670 & 0.0754 & 0.9680 \\
  $\beta_{8}$ & 0.0472 & 0.0521 & 0.9660 & 0.0687 & 0.0755 & 0.9720 \\
  $\beta_{9}$ & 0.0457 & 0.0521 & 0.9720 & 0.0662 & 0.0744 & 0.9560 \\
  $\beta_{10}$ & 0.0142 & 0.0186 & 0.9940 & 0.0350 & 0.0407 & 0.9880 \\
   \hline

\hline
\end{tabular}
\end{center}
\end{table}

\section{Application to Developing EHR Based Obesity Genetic Risk Prediction Model}

As one of the most serious public health problems in the 21st century, obesity affects about 12\% of adults globally \citep{gbd2017health}. Although there are lifestyle changes that can be made to prevent it, there is a strong genetic component to obesity, which is a risk factor for many other conditions such as cardiovascular disease and diabetes. We applied our proposed method to develop a genetic risk prediction model for obesity using data from Partner's Healthcare Biobank (PHB) where both genetic and EHR data are available for 30,685 participating patients.

Among the PHB subjets, 268 patients have their obesity status $\delta$ annotated by domain experts via manual chart review.  The genetic risk score (GRS) for obesity was constructed based on published log odds ratio information on 55 SNPs previously identified as significantly associated with obesity \citep{speliotes2010association,hung2015genetic}. For this analysis, we use patient age as time scale. We use 2 surrogates for event time: age at the first diagnostic code for obesity and age at the first NLP mention of obesity. In addition to the GRS, we included sex, and the first 5 principal components of 128 SNPs associated with the ancestry informative genetic markers \citep{Kosoy2009ancestry} to adjust for population stratification. We let $g(\cdot)$ be the logistic link and hence $\bbeta$ corresponds to log odds ratio of the risk factors.

The point estimators of $\bbeta_0$, as well as their 95\% confidence intervals are shown in Table \ref{tab-realdata}.
The results show that the supervised and semi-supervised point estimators  are reasonably consistent with each other.
Our analysis confirms that higher GRS is significantly associated with an elevated risks of
developing obesity, with estimated effect 0.610 and p-value 0.022 from the semi-supervised
method.
The negative association between male gender and risk of obesity in our analysis,
with estimated effect -0.365 from the semi-supervised
method,
is also consistent with the findings among the literatures \citep{WangBeydoun07}.
Moreover, the 95\% CIs from the semi-supervised estimator are always smaller than the supervised estimator.
For example, the estimated standard error of GRS coefficient is 0.327 in the supervised
estimator and 0.274 in the semi-supervised estimator, resulting in an efficiency gain of 1.415.
This again demonstrates the benefit of leveraging the information from the mis-measured
event times in our proposed procedure.

\begin{table}[ht]
\begin{center}
\caption{Point estimates and 95\% CIs of the risk prediction potential of sex, the obesity GRS and the first 5 PCs on
developing obesity, along with p-values from inverting the CIs.
} \label{tab-realdata}
\begin{tabular}{l|llcl|llcl}
\hline
\hline
& \multicolumn{4}{c|}{$\beta_\delta$} & \multicolumn{4}{c}{$\beta_{SSL}$} \\
\hline
& Est & SE & 95\% CI & PVal & Est & SE & 95\% CI & PVal \\
  \hline
 Male & -0.311 & 0.604 & (-1.670,0.739) & 0.354 & -0.365 & 0.568 & (-1.790,0.521) & 0.136 \\
  GRS &  0.572 & 0.327 & ( 0.071,1.334) & 0.038 &  0.610 & 0.274 & ( 0.081,1.167) & 0.022 \\
  PC1 &  0.946 & 0.337 & ( 0.321,1.671) & 0.006 &  0.922 & 0.318 & ( 0.306,1.588) & 0.006  \\
  PC2 &  0.692 & 0.587 & (-0.548,1.682) & 0.284 &  0.539 & 0.465 & (-0.548,1.340) & 0.208 \\
  PC3 &  0.107 & 0.447 & (-0.942,0.853) & 1.000 &  0.074 & 0.410 & (-0.794,0.864) & 0.838 \\
  PC4 &  0.495 & 0.328 & (-0.199,1.101) & 0.152 &  0.464 & 0.316 & (-0.211,1.052) & 0.152 \\
  PC5 &  0.327 & 0.336 & (-0.345,0.966) & 0.324 &  0.218 & 0.309 & (-0.404,0.793) & 0.470  \\
   \hline
\hline
\end{tabular}
\end{center}
\end{table}

\section{Discussion}

We proposed a robust SS estimators  for risk prediction modeling in the EHR setting. Our proposed SSL estimator is able to effectively integrate two sets of imperfect information on the survival time in the EHR database: (i) widely available but noisy surrogate event times; and (ii) current status information manually annotated for a limited set of patients. The SSL method can efficiently estimate the risk model without requiring precise information on the event time for any patients in the EHR, which greatly improves the feasibility of performing risk prediction modeling using noisy EHR data.  Our numerical results demonstrate that the SSL approach can significantly improve the efficiency of the estimation with large unlabelled data compared to the supervised estimator that only uses the current status data.

In the obesity risk modeling example with PHB data, we used age as the time scale and hence baseline is defined at birth. In such a case, only time invariant covariates such as sex and genetic information can be used as risk factors. However, our method is not restricted to such settings provided that a valid baseline can be defined and subjects are free of the event of interest at baseline.
For example, the baseline can be set as a year after the first encounter with the EHR. Patients without any ICD code and/or NLP mention of the disease of interest can be considered as free of event at baseline since ICD and NLP mentions are often highly sensitive but not specific \citep{liao2019high}. For such cases, standard risk factors beyond genetics such as lifestyle information can be included as covariates.




\bibliographystyle{chicago}
\bibliography{ref}

\begin{thebibliography}{}

\bibitem[\protect\citeauthoryear{Betensky, Rabinowitz, and Tsiatis}{Betensky
  et~al.}{2001}]{betensky2001computationally}
Betensky, R.~A., D.~Rabinowitz, and A.~A. Tsiatis (2001).
\newblock Computationally simple accelerated failure time regression for
  interval censored data.
\newblock {\em Biometrika\/}~{\em 88\/}(3), 703--711.

\bibitem[\protect\citeauthoryear{Braun, Gorfine, Katki, Ziogas, and
  Parmigiani}{Braun et~al.}{2018}]{braunnonparametric}
Braun, D., M.~Gorfine, H.~A. Katki, A.~Ziogas, and G.~Parmigiani (2018).
\newblock Nonparametric adjustment for measurement error in time to event data.
\newblock {\em Journal of the American Statistical Association\/}~{\em 113},
  11--25.

\bibitem[\protect\citeauthoryear{Cai and Cheng}{Cai and
  Cheng}{2007}]{CaiCheng08}
Cai, T. and S.~Cheng (2007, 12).
\newblock {Robust combination of multiple diagnostic tests for classifying
  censored event times}.
\newblock {\em Biostatistics\/}~{\em 9\/}(2), 216--233.

\bibitem[\protect\citeauthoryear{Calvert, Mao, Hoffman, Jay, Desautels,
  Mohamadlou, Chettipally, and Das}{Calvert et~al.}{2016}]{calvert2016using}
Calvert, J., Q.~Mao, J.~L. Hoffman, M.~Jay, T.~Desautels, H.~Mohamadlou,
  U.~Chettipally, and R.~Das (2016).
\newblock Using electronic health record collected clinical variables to
  predict medical intensive care unit mortality.
\newblock {\em Annals of medicine and surgery\/}~{\em 11}, 52--57.

\bibitem[\protect\citeauthoryear{Carroll, Fan, Gijbels, and Wand}{Carroll
  et~al.}{1997}]{CarrollEtal97}
Carroll, R.~J., J.~Fan, I.~Gijbels, and M.~P. Wand (1997).
\newblock Generalized partially linear single-index models.
\newblock {\em Journal of the American Statistical Association\/}~{\em
  92\/}(438), 477--489.

\bibitem[\protect\citeauthoryear{Chen and Sun}{Chen and
  Sun}{2010}]{chen2010multiple}
Chen, L. and J.~Sun (2010).
\newblock A multiple imputation approach to the analysis of interval-censored
  failure time data with the additive hazards model.
\newblock {\em Computational statistics \& data analysis\/}~{\em 54\/}(4),
  1109--1116.

\bibitem[\protect\citeauthoryear{Dvoretzky, Kiefer, and Wolfowitz}{Dvoretzky
  et~al.}{1956}]{DKW56}
Dvoretzky, A., J.~Kiefer, and J.~Wolfowitz (1956).
\newblock Asymptotic minimax character of the sample distribution function and
  of the classical multinomial estimator.
\newblock ~{\em 27\/}(3), 642--669.

\bibitem[\protect\citeauthoryear{Eapen, Liang, Fonarow, Heidenreich, Curtis,
  Peterson, and Hernandez}{Eapen et~al.}{2013}]{eapen2013validated}
Eapen, Z.~J., L.~Liang, G.~C. Fonarow, P.~A. Heidenreich, L.~H. Curtis, E.~D.
  Peterson, and A.~F. Hernandez (2013).
\newblock Validated, electronic health record deployable prediction models for
  assessing patient risk of 30-day rehospitalization and mortality in older
  heart failure patients.
\newblock {\em JACC: Heart Failure\/}~{\em 1\/}(3), 245--251.

\bibitem[\protect\citeauthoryear{{GBD 2015 Obesity Collaborators}}{{GBD 2015
  Obesity Collaborators}}{2017}]{gbd2017health}
{GBD 2015 Obesity Collaborators} (2017).
\newblock Health effects of overweight and obesity in 195 countries over 25
  years.
\newblock {\em New England Journal of Medicine\/}~{\em 377\/}(1), 13--27.

\bibitem[\protect\citeauthoryear{Han}{Han}{1987}]{han1987non}
Han, A.~K. (1987).
\newblock Non-parametric analysis of a generalized regression model: the
  maximum rank correlation estimator.
\newblock {\em Journal of Econometrics\/}~{\em 35\/}(2-3), 303--316.

\bibitem[\protect\citeauthoryear{Huang et~al.}{Huang
  et~al.}{1996}]{huang1996efficient}
Huang, J. et~al. (1996).
\newblock Efficient estimation for the proportional hazards model with interval
  censoring.
\newblock {\em The Annals of Statistics\/}~{\em 24\/}(2), 540--568.

\bibitem[\protect\citeauthoryear{Huang and Rossini}{Huang and
  Rossini}{1997}]{huang1997sieve}
Huang, J. and A.~Rossini (1997).
\newblock Sieve estimation for the proportional-odds failure-time regression
  model with interval censoring.
\newblock {\em Journal of the American Statistical Association\/}~{\em
  92\/}(439), 960--967.

\bibitem[\protect\citeauthoryear{Hung, Breen, Czamara, Corre, Wolf, Kloiber,
  Bergmann, Craddock, Gill, Holsboer, et~al.}{Hung
  et~al.}{2015}]{hung2015genetic}
Hung, C.-F., G.~Breen, D.~Czamara, T.~Corre, C.~Wolf, S.~Kloiber, S.~Bergmann,
  N.~Craddock, M.~Gill, F.~Holsboer, et~al. (2015).
\newblock A genetic risk score combining 32 snps is associated with body mass
  index and improves obesity prediction in people with major depressive
  disorder.
\newblock {\em BMC medicine\/}~{\em 13\/}(1), 86.

\bibitem[\protect\citeauthoryear{Jensen, Jensen, and Brunak}{Jensen
  et~al.}{2012}]{jensen2012mining}
Jensen, P.~B., L.~J. Jensen, and S.~Brunak (2012).
\newblock Mining electronic health records: towards better research
  applications and clinical care.
\newblock {\em Nature Reviews Genetics\/}~{\em 13\/}(6), 395.

\bibitem[\protect\citeauthoryear{Jin, Che, Liu, Zhang, Yin, and Wei}{Jin
  et~al.}{2018}]{jin2018predicting}
Jin, B., C.~Che, Z.~Liu, S.~Zhang, X.~Yin, and X.~Wei (2018).
\newblock Predicting the risk of heart failure with ehr sequential data
  modeling.
\newblock {\em Ieee Access\/}~{\em 6}, 9256--9261.

\bibitem[\protect\citeauthoryear{Jin, Ying, and Wei}{Jin
  et~al.}{2001}]{JinYingWei01}
Jin, Z., Z.~Ying, and L.~J. Wei (2001).
\newblock {A simple resampling method by perturbing the minimand}.
\newblock {\em Biometrika\/}~{\em 88\/}(2), 381--390.

\bibitem[\protect\citeauthoryear{Kosoy, Nassir, Tian, White, Butler, Silva,
  Kittles, Alarcon-Riquelme, Gregersen, Belmont, De~La~Vega, and Seldin}{Kosoy
  et~al.}{2009}]{Kosoy2009ancestry}
Kosoy, R., R.~Nassir, C.~Tian, P.~A. White, L.~M. Butler, G.~Silva, R.~Kittles,
  M.~E. Alarcon-Riquelme, P.~K. Gregersen, J.~W. Belmont, F.~M. De~La~Vega, and
  M.~F. Seldin (2009, Jan).
\newblock Ancestry informative marker sets for determining continental origin
  and admixture proportions in common populations in america.
\newblock {\em Human mutation\/}~{\em 30\/}(1), 69--78.
\newblock 18683858[pmid].

\bibitem[\protect\citeauthoryear{Liao, Sun, Cai, Link, Hong, Huang, Huffman,
  Gronsbell, Zhang, Ho, et~al.}{Liao et~al.}{2019}]{liao2019high}
Liao, K.~P., J.~Sun, T.~A. Cai, N.~Link, C.~Hong, J.~Huang, J.~E. Huffman,
  J.~Gronsbell, Y.~Zhang, Y.-L. Ho, et~al. (2019).
\newblock High-throughput multimodal automated phenotyping (map) with
  application to phewas.
\newblock {\em Journal of the American Medical Informatics Association,
  accepted\/}.

\bibitem[\protect\citeauthoryear{Lin, Oakes, and Ying}{Lin
  et~al.}{1998}]{lin1998additive}
Lin, D., D.~Oakes, and Z.~Ying (1998).
\newblock Additive hazards regression with current status data.
\newblock {\em Biometrika\/}~{\em 85\/}(2), 289--298.

\bibitem[\protect\citeauthoryear{Nolan and Pollard}{Nolan and
  Pollard}{1987}]{nolan1987}
Nolan, D. and D.~Pollard (1987, 06).
\newblock $u$-processes: Rates of convergence.
\newblock {\em Ann. Statist.\/}~{\em 15\/}(2), 780--799.

\bibitem[\protect\citeauthoryear{Oh, Shepherd, Lumley, and Shaw}{Oh
  et~al.}{2018}]{oh2018considerations}
Oh, E.~J., B.~E. Shepherd, T.~Lumley, and P.~A. Shaw (2018).
\newblock Considerations for analysis of time-to-event outcomes measured with
  error: Bias and correction with simex.
\newblock {\em Statistics in medicine\/}~{\em 37\/}(8), 1276--1289.

\bibitem[\protect\citeauthoryear{Rossini and Tsiatis}{Rossini and
  Tsiatis}{1996}]{rossini1996semiparametric}
Rossini, A. and A.~Tsiatis (1996).
\newblock A semiparametric proportional odds regression model for the analysis
  of current status data.
\newblock {\em Journal of the American Statistical Association\/}~{\em
  91\/}(434), 713--721.

\bibitem[\protect\citeauthoryear{Sherman}{Sherman}{1993}]{sherman1993limiting}
Sherman, R.~P. (1993).
\newblock The limiting distribution of the maximum rank correlation estimator.
\newblock {\em Econometrica: Journal of the Econometric Society\/}, 123--137.

\bibitem[\protect\citeauthoryear{Speliotes, Willer, Berndt, Monda,
  Thorleifsson, Jackson, Allen, Lindgren, Luan, M{\"a}gi, et~al.}{Speliotes
  et~al.}{2010}]{speliotes2010association}
Speliotes, E.~K., C.~J. Willer, S.~I. Berndt, K.~L. Monda, G.~Thorleifsson,
  A.~U. Jackson, H.~L. Allen, C.~M. Lindgren, J.~Luan, R.~M{\"a}gi, et~al.
  (2010).
\newblock Association analyses of 249,796 individuals reveal 18 new loci
  associated with body mass index.
\newblock {\em Nature genetics\/}~{\em 42\/}(11), 937.

\bibitem[\protect\citeauthoryear{Sun and Sun}{Sun and
  Sun}{2005}]{sun2005semiparametric}
Sun, J. and L.~Sun (2005).
\newblock Semiparametric linear transformation models for current status data.
\newblock {\em Canadian Journal of Statistics\/}~{\em 33\/}(1), 85--96.

\bibitem[\protect\citeauthoryear{Tian and Cai}{Tian and
  Cai}{2006}]{tian2006accelerated}
Tian, L. and T.~Cai (2006).
\newblock On the accelerated failure time model for current status and interval
  censored data.
\newblock {\em Biometrika\/}~{\em 93\/}(2), 329--342.

\bibitem[\protect\citeauthoryear{Uno, Ritzwoller, Cronin, Carroll, Hornbrook,
  and Hassett}{Uno et~al.}{2018}]{uno2018determining}
Uno, H., D.~P. Ritzwoller, A.~M. Cronin, N.~M. Carroll, M.~C. Hornbrook, and
  M.~J. Hassett (2018).
\newblock Determining the time of cancer recurrence using claims or electronic
  medical record data.
\newblock {\em JCO clinical cancer informatics\/}~{\em 2}, 1--10.

\bibitem[\protect\citeauthoryear{Van Der~Laan and Robins}{Van Der~Laan and
  Robins}{1998}]{van1998locally}
Van Der~Laan, M.~J. and J.~M. Robins (1998).
\newblock Locally efficient estimation with current status data and
  time-dependent covariates.
\newblock {\em Journal of the American Statistical Association\/}~{\em
  93\/}(442), 693--701.

\bibitem[\protect\citeauthoryear{Wang and Beydoun}{Wang and
  Beydoun}{2007}]{WangBeydoun07}
Wang, Y. and M.~A. Beydoun (2007, 05).
\newblock {The Obesity Epidemic in the United States—Gender, Age,
  Socioeconomic, Racial/Ethnic, and Geographic Characteristics: A Systematic
  Review and Meta-Regression Analysis}.
\newblock {\em Epidemiologic Reviews\/}~{\em 29\/}(1), 6--28.

\end{thebibliography}

{\LARGE\bf Supplementary Material for ``Risk Prediction with Imperfect Survival Outcome Information from Electronic Health Records''}

\setcounter{section}{0}
\renewcommand\thesection{Appendix \Alph{section}}

\renewcommand\thefigure{A\arabic{figure}}
\renewcommand\theequation{A.\arabic{equation}}
\renewcommand\thetable{A\arabic{table}}
\renewcommand\thelemma{A\arabic{lemma}}
\renewcommand\thesubsection{\Alph{section}\arabic{subsection}}
\renewcommand\thedefinition{A\arabic{definition}}
\renewcommand\theremark{A\arabic{remark}}

\setcounter{equation}{0}
\setcounter{lemma}{0}

\section{Verification of Single Index Model}\label{app:sim}

Here we show that each of the $\Tscr_k$ follows a single index model.
Suppose $f_{\epsilon_k}(u)$ is the probability density function of $\varepsilon_{ki}$.
By direct calculation, we find that conditional distribution function of $\Tscr_{ki}$ is
\begin{eqnarray*}
P(\Tscr_{ki}\leq t|\bZ_i)&=&P(\Hsc(\Tscr_{ki})\leq \Hsc(t)|\bZ_i)=P(\Hsc(T_i)+\epsilon_{ki}\leq \Hsc(t)|\bZ_i)\\
&=&P(\Hsc(T)\leq \Hsc(t)-\epsilon_{ki}|\bZ_i)=P( T \leq \Hsc^{-1} ( \Hsc(t) -\epsilon_{ki})|\bZ_i)\\
&=&\int g(h_{0}(\Hsc^{-1} ( \Hsc(t) -u))+\bbeta_{0}\trans\bZ_i)  f_{\epsilon_k}(u)d u,
\end{eqnarray*}
which is still a increasing function of $\bbeta_{0}\trans\bZ_i$ and $\mathbf{\mathcal{B}}_0\trans \bZ_i$.
The proof here shows that the truncated $\Tscr_{ki} \wedge \Cscr_r$ also follows the single index model,
as
$$
P(\Tscr_{ki}\wedge \Cscr_r \leq t|\bZ_i) =
\left\{
\begin{array}{cc}
P(\Tscr_{ki} \leq t|\bZ_i) & t < \Cscr_r, \\
1 & t \ge \Cscr_r
\end{array}
\right.
$$
is also an increasing function of $\bbeta_{0}\trans\bZ_i$.

\section{Asymptotic property of the initial estimator}\label{app:init}

Under the semiparametric single index model \eqref{eq:STM-moment} for current status $\delta_i$, our initial estimation is a special case of the quasi-likelihood approach considered in \cite{CarrollEtal97}.
To be specific, our initial estimator is the fully iterated estimator with misspecified variance in their paper.
By their Remark 2 and Theorem 3, we have the consistency and asymptotic normality of $\bbetahat\subdelta$.
In the following, we derive the asymptotic distribution of $\bbetahat\subdelta$, which is not stated in \cite{CarrollEtal97}.

The estimators $\widehat{h}$ and $\bbetahat\subdelta$ satisfy
\begin{gather}
 \sum_{i=1}^n  \bZ_i[\delta_i - g\{\widehat{h}(C_i) + \bbetahat\subdelta\trans\bZ_i\}] = 0 \notag \\
 \sum_{i=1}^n  \Kbb_h(C_i - t) [\delta_i - g\{\widehat{h}(t) + \bbetahat\subdelta\trans\bZ_i\}] = 0, \;
 t = C_1, C_2, \dots, C_n.  \label{ee:init}
\end{gather}

We denote the remainder rate as
$$
R_n = \sqrt{n}\left\{\left\|\widehat{h}-h_0\right\|_\infty+\left\|\bbetahat\subdelta-\bbeta_0\right\|_2\right\}
\left\{\left\|\widehat{h}-h_0\right\|_\infty+\left\|\bbetahat\subdelta-\bbeta_0\right\|_2+(nh)^{-1/2}+h^2\right\}
+ \sqrt{n}h^2.
$$

To derive the asymptotic distribution of $\bbetahat\subdelta$,
we first consider the local approximation of the first sets of equations in \eqref{ee:init},
\begin{align*}
  n^{-1/2} \sum_{i=1}^n  \bZ_i[\delta_i - g\{h_0(C_i) + \bbeta_0\trans\bZ_i\}]
 = & n^{-1/2} \sum_{i=1}^n  \bZ_i[ g\{\widehat{h}(C_i) + \bbetahat\subdelta\trans\bZ_i\} - g\{h_0(C_i) + \bbeta_0\trans\bZ_i\}] \\
 = & n^{-1/2} \sum_{i=1}^n  \bZ_i g'\{h_0(C_i) + \bbeta_0\trans\bZ_i\} \{ \widehat{h}(C_i)  - h_0(C_i)\} \\
  & + \sqrt{n} \E\left[ \bZ_i\bZ_i\trans g'\{h_0(C_i) + \bbeta_0\trans\bZ_i\}\right] \{  \bbetahat\subdelta - \bbeta_0\} \\
 & + O_p\left(R_n\right).
\end{align*}
The key is the analysis of the first order term for the nonparametric component
\begin{equation}\label{eq:1st-h}
  n^{-1/2} \sum_{i=1}^n  \bZ_i g'\{h_0(C_i) + \bbeta_0\trans\bZ_i\} \{ \widehat{h}(C_i)  - h_0(C_i)\}
\end{equation}
From the second set of equations in \eqref{ee:init}, we have
\begin{align*}
& n^{-1} \sum_{i=1}^n  \Kbb_h(C_i - C_j) [\delta_i - g\{h_0(C_j) + \bbeta_0\trans\bZ_i\}] \\
= & n^{-1} \sum_{i=1}^n  \Kbb_h(C_i - C_j) [ g\{\widehat{h}(C_j) + \bbetahat\subdelta\trans\bZ_i\} - g\{h_0(C_j) + \bbeta_0\trans\bZ_i\}] \\
= & n^{-1} \sum_{i=1}^n  \Kbb_h(C_i - C_j) g'\{h_0(C_j) + \bbeta_0\trans\bZ_i\} \{ \widehat{h}(C_j)  - h_0(C_j)\} \\
& + n^{-1} \sum_{i=1}^n \bZ_i \Kbb_h(C_i - C_j) g'\{h_0(C_j) + \bbeta_0\trans\bZ_i\}\{  \bbetahat\subdelta - \bbeta_0\} \\
& + O_p\left(n^{-1/2}R_n\right) .
\end{align*}
With $f_c(t)$ being the density of censoring time $C_i$ and
$$
  a_0(t) = \E[f_c(t) g'\{h_0(t) + \bbeta_0\trans\bZ_i\}]
  , \;
    \ba_1(t) = \E[f_c(t) g'\{h_0(t) + \bbeta_0\trans\bZ_i\}\bZ_i]
    .
$$
defined in \eqref{def:BSigd},
we may approximate the estimation error of the nonparametric component $\widehat{h}(C_i)  - h_0(C_i)$ as
\begin{align}
 & \widehat{h}(C_j)  - h_0(C_j) \notag \\
 =  & \left\{a_0(C_j)\right\}^{-1}\left(
   (nh)^{-1} \sum_{i=1}^n  \Kbb_h(C_i - C_j) [\delta_i - g\{h_0(C_j) + \bbeta_0\trans\bZ_i\}]
 -
   \ba_1(C_j)\trans \{  \bbetahat\subdelta - \bbeta_0\} \right)\notag \\
&  + O_p\left(n^{-1/2}R_n\right).
\label{eq:h-infl}
\end{align}
Here we also obtain
$$
\|\widehat{h}-h_0\|_\infty = O_p\left((nh)^{-1/2}+h^2 + \|\bbetahat\subdelta-\bbeta_0\|_2\right),
$$
so we have an upper bound for the remainder rate
$$
R_n = O_p\left(\sqrt{n}
\left\{\left\|\bbetahat\subdelta-\bbeta_0\right\|_2+(nh)^{-1/2}+h^2\right\}^2
+ \sqrt{n}h^2.\right).
$$
Plugging \eqref{eq:h-infl} to \eqref{eq:1st-h}, we have
\begin{align*}
  & n^{-1/2} \sum_{i=1}^n  \bZ_i g'\{h_0(C_i) + \bbeta_0\trans\bZ_i\} \{ \widehat{h}(C_i)  - h_0(C_i)\} \\
  = & n^{-1/2} \sum_{i=1}^n  \bZ_i g'\{h_0(C_i) + \bbeta_0\trans\bZ_i\} \left\{a_0(C_i)\right\}^{-1}
   n^{-1} \sum_{j=1}^n  \Kbb_h(C_i - C_j) [\delta_j - g\{h_0(C_i) + \bbeta_0\trans\bZ_j\}] \\
   & - \sqrt{n}  \E\left[ \bZ_i\ba_1(C_i) g'\{h_0(C_i) + \bbeta_0\trans\bZ_i\} /a_0(C_i)\right]
\trans \{  \bbetahat\subdelta - \bbeta_0\}  \\
  & + O_p(R_n).
\end{align*}
Let
$
\bb(t) =\ba_1(t)/a_0(t)
$
as defined in \eqref{def:BSigd}.
By change the order of the sum over $i$ and $j$, we have
\begin{align*}
& n^{-1/2} \sum_{i=1}^n  \bZ_i g'\{h_0(C_i) + \bbeta_0\trans\bZ_i\} \left\{a_0(C_i)\right\}^{-1}
   n^{-1} \sum_{j=1}^n  \Kbb_h(C_i - C_j) [\delta_j - g\{h_0(C_i) + \bbeta_0\trans\bZ_j\}] \\
 =  & n^{-1/2}\sum_{j=1}^n[\delta_j - g\{h_0(C_j) + \bbeta_0\trans\bZ_j\}] n^{-1}\sum_{i=1}^n \bZ_i g'\{h_0(C_i) + \bbeta_0\trans\bZ_i\} \left\{a_0(C_i)\right\}^{-1} \Kbb_h(C_i - C_j) \\
& - n^{-1/2}\sum_{j=1}^n n^{-1}\sum_{i=1}^n \bZ_i g'\{h_0(C_i) + \bbeta_0\trans\bZ_i\} \left\{a_0(C_i)\right\}^{-1} \Kbb_h(C_i - C_j) \\
& \qquad \times[g\{h_0(C_j) + \bbeta_0\trans\bZ_j\} - g\{h_0(C_i) + \bbeta_0\trans\bZ_j\}] \\
 = &  n^{-1/2}\sum_{j=1}^n \bb(C_j) [\delta_j - g\{h_0(C_j) + \bbeta_0\trans\bZ_j\}]
 + O_p(R_n).
\end{align*}
Gathering the results, we have reached
\begin{align*}
& \sqrt{n}\{  \bbetahat\subdelta - \bbeta_0\}\trans
\E\left[ \left\{\bZ_i - \ba_1(C_i)/a_0(C_i)\right\}\bZ_i\trans g'\{h_0(C_i) + \bbeta_0\trans\bZ_i\}\right]  \\
= &  n^{-1/2} \sum_{i=1}^n  \{\bZ_i-\bb(C_i)\}[\delta_i - g\{h_0(C_i) + \bbeta_0\trans\bZ_i\}] + O_p(R_n).
\end{align*}
As long as $n^{-1/2} \ll h \ll n^{-1/4}$,
we may deduce from the representation the root-n consistency of $\bbetahat\subdelta$,
$$
\|\bbetahat\subdelta - \bbeta_0\|_2 = O_p\left(\sqrt{n}\right), \; R_n = o_p(1).
$$
Using the notations $\Bbb$ and $\Sigbb\subdelta$ defined in \eqref{def:BSigd},
we may express the asymptotic distribution of $\bbeta\subdelta$ as
\begin{align}
\sqrt{n}\{  \bbetahat\subdelta - \bbeta_0\} = & n^{-1/2} \sum_{i=1}^n  \Bbb^{-1}\{\bZ_i-\bb(C_i)\}[\delta_i - g\{h_0(C_i) + \bbeta_0\trans\bZ_i\}] + o_p(1) \notag \\
\leadsto & N\left(\mathbf{0}, \Bbb^{-1} \Sigbb\subdelta(\Bbb^{-1})\trans\right)
\tag{\ref{eq:init_asymp}}
\end{align}
as stated in the main text.

\section{Asymptotic property of rank correlation}\label{app:Sk}
We focus our analysis on the numerator of $\bShat_k(\bBsc)$,
$$
\bShat_k^*(\bBsc) = N^{-2} \sum_{i=1}^N\sum_{j = 1}^N(\bZ_i - \bZ_j)\Kbb_{h'}(\bBsc\trans\bZ_i - \bBsc\trans\bZ_j)\frac{I(\Xscr_{ki} < \Xscr_{kj})\Delta_{ki}}{\widehat{G}(\Xscr_{ki})^2}.
$$
The denominator of $\bShat_k(\bBsc)$ does not contain the parameter $\bBsc$,
and it is easy to show its convergence to its mean,
$$
N^{-2}\sum_{i=1}^N\sum_{j = 1}^N\frac{I(\Xscr_{ki} < \Xscr_{kj})\Delta_{ki}}{\widehat{G}(\Xscr_{ki})^2}
\to \E\left\{\frac{I(\Xscr_{ki} < \Xscr_{kj})\Delta_{ki}}{G(\Xscr_{ki})^2}\right\} = 1/2.
$$
We may eventually substitute the denominator by its limit $1/2$ through a Slutsky's Theorem argument,
\begin{equation}\label{eq:Shat*}
2\bShat_k^*(\bBsc) - \bShat_k(\bBsc)
= o_p\left(\bShat_k^*(\bBsc_0) + \|\bBsc - \bBsc_0\|_2\right).
\end{equation}

Now, we establish the connection between $\bShat_k^*(\bBsc)$
and the rank correlation
\begin{equation}\tag{\ref{def:Qk}}
   \widehat{Q}_k(\bBsc) =  N^{-2} \sum_{i=1}^N\sum_{j = 1}^N I(\bBsc\trans\bZ_i - \bBsc\trans\bZ_j > 0)\frac{I(\Xscr_{ki} < \Xscr_{kj})\Delta_{ki}}{\widehat{G}(\Xscr_{ki})^2}.
\end{equation}
We define the anti-derivative of the symmetric smooth kernel $\Kbb(\cdot)$ as
\begin{equation}
\Fbb(x) = \int_{-\infty}^x \Kbb(u) du, \;
  \Fbb_{h'}(x) = \int_{-\infty}^x \Kbb_{h'}(u) du = \Fbb(x/h').
\end{equation}
Assuming $\Kbb(\cdot)$ is a symmetric probability density function with second moment,
then functions $\Fbb(x) - I(x>0)$ and $x\{\Fbb(x) - I(x>0)\}$ are absolutely integrable
over $\Rbb$,
and
$$
\int_{-\infty}^{\infty} \Fbb(x) - I(x>0) dx =0.
$$
Thus for any continuously differentiable function $f(x)$, we have
\begin{equation}\label{eq:Fbb}
  \int_{-\infty}^{\infty} \{\Fbb_{h'}(x) - I(x>0)\}f(x) dx
  = O(h'f'(0)).
\end{equation}
Our score $\bShat_k^*(\bBsc)$ is the gradient of the smoothed rank correlation
\begin{equation}
  \widehat{Q}_k^*(\bBsc) =  N^{-2} \sum_{i=1}^N\sum_{j = 1}^N\Fbb_{h'}(\bBsc\trans\bZ_i - \bBsc\trans\bZ_j)\frac{I(\Xscr_{ki} < \Xscr_{kj})\Delta_{ki}}{\widehat{G}(\Xscr_{ki})^2}, \;
  \bShat_k^*(\bBsc) = \nabla \widehat{Q}_k^*(\bBsc).
\end{equation}
\cite{CaiCheng08} has thoroughly studied the rank correlation without smoothing $\widehat{Q}_k(\bBsc)$,
and
we shall develop the properties for $\widehat{Q}_k^*(\bBsc)$ by connecting it to those of $\widehat{Q}_k(\bBsc)$.

By the uniform consistency of empirical distribution $\widehat{G}(\cdot)$ \citep{DKW56},
we may replace the estimated $\widehat{G}(\cdot)$ by the true $G(\cdot)$ in $\widehat{Q}_k^*(\bBsc)$
and $\widehat{Q}_k(\bBsc)$ with an $O_p\left(N^{-1/2}\right)$ error,
\begin{gather}
  \widetilde{Q}_k^*(\bBsc) =  N^{-2} \sum_{i=1}^N\sum_{j = 1}^N\Fbb_{h'}(\bBsc\trans\bZ_i - \bBsc\trans\bZ_j)\frac{I(\Xscr_{ki} < \Xscr_{kj})\Delta_{ki}}{G(\Xscr_{ki})^2}
  = \widehat{Q}_k^*(\bBsc) + O_p\left(N^{-1/2}\right), \notag \\
  \widetilde{Q}_k(\bBsc) =  N^{-2} \sum_{i=1}^N\sum_{j = 1}^NI(\bBsc\trans\bZ_i - \bBsc\trans\bZ_j > 0)\frac{I(\Xscr_{ki} < \Xscr_{kj})\Delta_{ki}}{G(\Xscr_{ki})^2}
  = \widehat{Q}_k(\bBsc) + O_p\left(N^{-1/2}\right).
\end{gather}
Define the limiting processes for $\widetilde{Q}_k^*(\bBsc)$
and $\widetilde{Q}_k(\bBsc)$ as
 $q_k(\bBsc) = \E\{q_{kij}(\bBsc)\}$,
 $q_k^*(\bBsc) = \E\{q_{kij}^*(\bBsc)\}$ with
\begin{align}
q_{ijk}(\bBsc) = \frac{1}{2}& \left\{I(\bBsc\trans\bZ_i - \bBsc\trans\bZ_j > 0)\frac{I(\Xscr_{ki} < \Xscr_{kj})\Delta_{ki}}{G(\Xscr_{ki})^2} \right. \notag \\
& \left.\quad + I(\bBsc\trans\bZ_j - \bBsc\trans\bZ_i > 0)\frac{I(\Xscr_{kj} < \Xscr_{ki})\Delta_{kj}}{G(\Xscr_{kj})^2}\right\}, \notag \\
  q_{ijk}^*(\bBsc) =  \frac{1}{2}& \left\{\Fbb_{h'}(\bBsc\trans\bZ_i - \bBsc\trans\bZ_j)\frac{I(\Xscr_{ki} < \Xscr_{kj})\Delta_{ki}}{G(\Xscr_{ki})^2} \right. \notag\\
  & \left.\quad + \Fbb_{h'}(\bBsc\trans\bZ_j - \bBsc\trans\bZ_i)\frac{I(\Xscr_{kj} < \Xscr_{ki})\Delta_{kj}}{G(\Xscr_{kj})^2}\right\}.
\end{align}
We assume that $q_k(\bBsc)$ is twice continuously differentiable.
By a U process argument \citep{nolan1987}, we have
$$
\sup_{\bBsc: \|\bBsc\|_2\le 2} |\widetilde{Q}_k(\bBsc) - q(\bBsc)| + |\widetilde{Q}^*_k(\bBsc) - q^*(\bBsc)| \to 0
$$
almost surely on the compact support $\|\bBsc\|_2\le 2$.
By the property of $\Fbb_{h'}$ as in \eqref{eq:Fbb},
we have
$$
\sup_{\bBsc: \|\bBsc\|_2\le 2} | q^*(\bBsc) -  q(\bBsc)|
= O(h') = o(1).
$$
Thus, we have
\begin{equation}\label{eq:lim_Q}
\sup_{\bBsc: \|\bBsc\|_2\le 2} |\widetilde{Q}^*_k(\bBsc) - q(\bBsc)| \to 0, \; \text{almost surely.}
\end{equation}

Next, we show that $\widetilde{\bS}_k^*$ has a continuous limit
so that
 we may establish its limit as $\dot{\bq}_k$.
We first consider
\begin{align}
\widetilde{\bS}_k^*(\bBsc) = & N^{-2} \sum_{i=1}^N\sum_{j = 1}^N \dot{\bq}_{ijk}^*(\bBsc), \notag \\
\dot{\bq}_{ijk}^*(\bBsc) = &  \frac{1}{2}\left\{
(\bZ_i - \bZ_j)\Kbb_{h'}(\bBsc\trans\bZ_i - \bBsc\trans\bZ_j)\frac{I(\Xscr_{ki} < \Xscr_{kj})\Delta_{ki}}{G(\Xscr_{ki})^2} \right. \\
& + \left. (\bZ_j - \bZ_i)\Kbb_{h'}(\bBsc\trans\bZ_j - \bBsc\trans\bZ_i)\frac{I(\Xscr_{kj} < \Xscr_{ki})\Delta_{kj}}{G(\Xscr_{kj})^2}
\right\}
\end{align}
Using the same U process argument \citep{nolan1987}, we can show that for $h' \gg 1/N$
\begin{equation}\label{eq:Sk-qk}
\sup_{\bBsc: \|\bBsc\|_2\le 2}\|\widetilde{\bS}_k^*(\bBsc)-\dot{\bq}_k^*(\bBsc)\|_2 \to 0, \; a.s., \quad
\dot{\bq}_k^* = \E\{\dot{\bq}_{ijk}^*(\bBsc)\} = \nabla q_k^*(\bBsc).
\end{equation}
where we use the dot notation for gradient.
As the smoothed version of $\dot{\bq}_k$,
the continuity of $\dot{\bq}^*_k$ is implied by the continuity of $\dot{\bq}_k$.
Thus, we have shown that $\widehat{\bS}_k^*(\bBsc)$ converges to a continuous limit.
By a calculus result, the differentiation and limit are commutable for a uniformly convergent function sequence
whose derivative converges uniformly to a continuous function.
Combining \eqref{eq:lim_Q} and \eqref{eq:Sk-qk},
we apply the calculus result point-wisely in the probability space to obtain
\begin{equation}\label{eq:lim_Sk}
 \sup_{\bBsc: \|\bBsc\|_2\le 2}\|\widetilde{\bS}_k^*(\bBsc)-\dot{\bq}_k(\bBsc)\|_2 \to 0, \; a.s., \quad
\dot{\bq}_k = \nabla q_k(\bBsc).
\end{equation}

Repeating the local quadratic expansion in \cite{CaiCheng08} from the arguments originally in \cite{sherman1993limiting},
we have for $\|\bBsc - \bBsc_0\|_2 = O(N^{-1/2})$,
\begin{equation}\label{eq:Q-quad}
  \widetilde{Q}_k(\bBsc)- \widetilde{Q}_k(\bBsc_0)
  = \frac{1}{2}(\bBsc - \bBsc_0)\trans \Abb_k (\bBsc - \bBsc_0)
  + \frac{1}{N}\sum_{i=1}^N \dot{\bq}_{k,i}(\bBsc_0)\trans (\bBsc - \bBsc_0)
  + o_p(N^{-1}+\|\bBsc - \bBsc_0\|_2^2)
\end{equation}
with $\Abb_k$ and $\dot{\bq}_{k,i}$ defined in \eqref{def:qki_Ak_qkc}.
Through the same arugment, we establish for $\|\bBsc - \bBsc_0\|_2 = O(n^{-1/2})$
and $h' = o(N^{-1/4})$,
\begin{equation}\label{eq:Stil_approx}
  \widetilde{\bS}_k^*(\bBsc)
  = \Abb_k (\bBsc - \bBsc_0)
  + \frac{1}{N}\sum_{i=1}^N \dot{\bq}_{k,i}(\bBsc_0)
  + o_p(N^{-1/2}+\|\bBsc - \bBsc_0\|_2).
\end{equation}

Now, we study the effect of estimated $\widehat{G}$ on the asymptotic distribution of $\bShat_k^*$.
Using the rate of consistency $\|\widehat{G}-G\|_\infty = O_p\left(N^{-1/2}\right)$ \citep{DKW56},
we may approximate the following error
\begin{align}
& \widehat{\bS}_k^*(\bBsc) -  \widetilde{\bS}_k^*(\bBsc) \notag \\
= &
N^{-2} \sum_{i=1}^N\sum_{j = 1}^N(\bZ_i - \bZ_j)\Kbb_{h'}(\bBsc\trans\bZ_i - \bBsc\trans\bZ_j)\frac{I(\Xscr_{ki} < \Xscr_{kj})\Delta_{ki}}{G(\Xscr_{ki})^3}
\{\widehat{G}(\Xscr_{ki}) - G(\Xscr_{ki})\}  + O_p\left(N^{-1}\right) \notag\\
= & N^{-1}\sum_{l=1}^N N^{-2}\sum_{i\neq l}\sum_{j \neq l} (\bZ_i - \bZ_j)\Kbb_{h'}(\bBsc\trans\bZ_i - \bBsc\trans\bZ_j)\frac{I(\Xscr_{ki} < \Xscr_{kj})\Delta_{ki}}{G(\Xscr_{ki})^3}
\{I(C_l \ge \Xscr_{ki}) - G(\Xscr_{ki})\}  \notag \\
& + O_p\left(N^{-1}\right).
\end{align}
To use the U-process argument, we symmetrize the terms in $\widehat{\bS}_k^*(\bBsc) -  \widetilde{\bS}_k^*(\bBsc)$,
\begin{align}
 \dot{\bq}^{c*}_{ijk}(c,\bBsc)  = & \frac{1}{2}\left[(\bZ_i - \bZ_j)\Kbb_{h'}(\bBsc\trans\bZ_i - \bBsc\trans\bZ_j)\frac{I(\Xscr_{ki} < \Xscr_{kj})\Delta_{ki}}{G(\Xscr_{ki})^3}
\{I(c \ge \Xscr_{ki}) - G(\Xscr_{ki})\} \right. \notag \\
& \quad \left.+ (\bZ_j - \bZ_i)\Kbb_{h'}(\bBsc\trans\bZ_j - \bBsc\trans\bZ_i)\frac{I(\Xscr_{kj} < \Xscr_{ki})\Delta_{kj}}{G(\Xscr_{kj})^3}
\{I(c \ge \Xscr_{kj}) - G(\Xscr_{kj})\} \right]
\end{align}
and its sample average and expectation as
\begin{align}\label{}
\dot{\widehat{\bq}}^{c*}_{k}(c,\bBsc) = N^{-2}\sum_{i\neq l}\sum_{j \neq l} \dot{\bq}^{c*}_{ijk}(c,\bBsc) , \; \dot{\bq}^{c*}_{k}(c,\bBsc) = \E \{\dot{\bq}^{c*}_{ijk}(c,\bBsc)\}.
\end{align}
When $h' \gg N^{-1}$, we apply the U-process argument \citep{nolan1987}
$$
\sup_{\bBsc: \|\bBsc\|_2\le 2}\sup_{c\in [0,\tau]}\left| \dot{\hat{\bq}}^{c*}_{k}(c, \bBsc) - \dot{\bq}^{c*}_{k}(c, \bBsc)\right|
= O_p\left(N^{-1/2}\right),
$$
where $\tau$ is the maximal censoring time.
Using the fact that
$$
\E\{\dot{\bq}^{c*}_{k}(C_l,\bBsc)\}
= \E\left[\E\{\dot{\widehat{\bq}}^{c*}_{ijk}(C_l,\bBsc)\mid C_l\}\right] = 0,
$$
we establish
\begin{align}
    \widehat{\bS}_k^*(\bBsc) -  \widetilde{\bS}_k^*(\bBsc)
  = & N^{-1}\sum_{l=1}^{N}\dot{\bq}^{c*}_{k}(C_l,\bBsc)
  +  N^{-1}\sum_{l=1}^{N}\left\{\dot{\bq}^{c*}_{k}(C_l,\bBsc) - \dot{\hat{\bq}}^{c*}_{k}(C_l,\bBsc)\right\}
  +  O_p\left(N^{-1}\right) \notag \\
  = & N^{-1}\sum_{l=1}^{N}\dot{\bq}^{c*}_{k}(C_l,\bBsc)+  O_p\left(N^{-1}\right)
  \label{eq:Shat-Stilde}
\end{align}
Combining \eqref{eq:Stil_approx} and \eqref{eq:Shat-Stilde}, we have shown

\begin{equation}\label{eq:Shat_approx_c*}
\widehat{\bS}^*_k(\bBsc) = N^{-1}\sum_{i=1}^{N} \left\{\dot{\bq}_{k,i}(\bBsc_0)+ \dot{\bq}^{c*}_k(C_i,\bBsc_0)\right\} + \Abb_k (\bBsc - \bBsc_0) + o_p(\|\bBsc - \bBsc_0\|_2 + N^{-1/2}).
\end{equation}
Repeating the argument from which we obtain \eqref{eq:lim_Sk},
we have
\begin{equation}
 \sup_{\bBsc: \|\bBsc\|_2\le 2}\sup_{c\in [0,\tau]} |\dot{\bq}^{c*}_{k}(c,\bBsc) - \dot{\bq}^{c}_{k}(c,\bBsc)| \to 0, \; a.s.
\end{equation}
whenever $\dot{\bq}^{c}_{k}$ defined in \eqref{def:qki_Ak_qkc} exists.
Since
$$
\E\{\dot{\bq}^{c*}_{k}(C_i,\bBsc)\} = \E\{\dot{\bq}^{c}_{k}(C_i,\bBsc)\} = \mathbf{0},
$$
we can substitute $\dot{\bq}^{c*}_{k}$ in \eqref{eq:Shat_approx_c*} by $\dot{\bq}^{c}_{k}$ with
an $o_p\left(N^{-1/2}\right)$ error,
\begin{equation}
\widehat{\bS}^*_k(\bBsc) = N^{-1}\sum_{i=1}^{N} \left\{\dot{\bq}_{k,i}(\bBsc_0)+ \dot{\bq}^{c}_k(C_i,\bBsc_0)\right\} + \Abb_k (\bBsc - \bBsc_0) + o_p(\|\bBsc - \bBsc_0\|_2 + N^{-1/2}).
\end{equation}
By \eqref{eq:Shat*}, we finally establish for $\|\bBsc - \bBsc_0\|_2 = O(n^{-1/2})$,
\begin{equation}\label{eq:Shat_approx}
\widehat{\bS}_k(\bBsc) = \frac{2}{N}\sum_{i=1}^{N} \left\{\dot{\bq}_{k,i}(\bBsc_0)+ \dot{\bq}^{c}_k(C_i,\bBsc_0)\right\} + 2\Abb_k (\bBsc - \bBsc_0) + o_p(\|\bBsc - \bBsc_0\|_2 + N^{-1/2}).
\end{equation}

By the analyses of $\bbetahat\subdelta$ and $\widehat{\bS}_k(\bBsc)$,
we have the local asymptotic approximation of $\widehat{\bS}(\bBschat\subdelta)$,
\begin{align}
 \widehat{\bS}(\bBschat\subdelta) = & \frac{2}{N}\sum_{i=1}^{N} \left\{\dot{\bq}_i(\bBsc_0)+ \dot{\bq}^c(C_i,\bBsc_0)\right\} +2\Abb (\bBschat\subdelta - \bBsc_0)
 + o_p(n^{-1/2} + N^{-1/2}) \notag \\
 = & \frac{2}{N}\sum_{i=1}^{N} \left\{\dot{\bq}_i(\bBsc_0)+ \dot{\bq}^c(C_i,\bBsc_0)\right\} + 2\Abb (\bbetahat\subdelta - \bbeta_0)/\|\bbeta_0\|_2 \notag \\
& + 2\Abb \bbeta_0 (\|\bbetahat\subdelta\|_2^{-1} - \|\bbeta_0\|_2^{-1})
 + o_p(n^{-1/2} + N^{-1/2}) \notag \\
 = & \frac{2}{n}\sum_{i=1}^{n}\Abb \|\bbeta_0\|_2^{-1} \Bbb^{-1}\{\bZ_i-\bb(C_j)\}[\delta_i - g\{h_0(C_i) + \bbeta_0\trans\bZ_i\}] \notag \\
 & +\frac{2}{N}\sum_{i=1}^{N} \left\{\dot{\bq}_i(\bBsc_0)+ \dot{\bq}^c(C_i,\bBsc_0)\right\} + o_p(n^{-1/2} + N^{-1/2}). \label{eq:Sk-bhat-detail}
\end{align}
Here we stack $\widehat{\bS}_k$, $\dot{\bq}_{k,i}$, $\dot{\bq}^c_k$ and $\Abb_k$
to produce $\widehat{\bS}$, $\dot{\bq}_{i}$, $\dot{\bq}^c$ and $\Abb$
as described in \eqref{def:qki_Ak_qkc}
when $K \ge 2$.

\section{Asymptotic Properties of the SSL Estimator}\label{app:SSL}

\subsection*{\underline{Asymptotic collinearity of $\widehat{\bS}(\bBschat\subdelta)$}}
We first show a property of the quantities derived from the rank correlation in \eqref{def:qki_Ak_qkc}.
Since $\widehat{Q}_k(\bBsc)$ is invariant to the scale of $\bBsc$, its limit $q_k(\bBsc)$ must also
be invariant to the scale of $\bBsc$,
$$
\widehat{Q}_k(\bBsc) = \widehat{Q}_k(c \bBsc), \;
q_k(\bBsc) = q_k(c\bBsc), \; \forall c>0.
$$
Consequently, we have
\begin{equation}\label{eq:qdot-b0}
  \dot{\bq}_k(\bBsc_0)\trans \bbeta_0 = \|\bbeta_0\|_2^{-1}\lim_{t \to 0} t^{-1} \{ q_k(\bBsc_0+t\bBsc_0) -  q_k(\bBsc_0)\}
   = 0.
\end{equation}
Following the same argument, we also have
\begin{equation}\label{eq:qcdot-Ak-b0}
  \dot{\bq}^c_k(C_i,\bBsc_0) \trans \bbeta_0 = 0, \;
  \Abb_k \bbeta_0 = \Abb_k\trans \bbeta_0 = \mathbf{0}.
\end{equation}
The properties \eqref{eq:qdot-b0} and \eqref{eq:qcdot-Ak-b0} later play an important
role in the asymptotic distribution of the semi-supervised estimator $\bbetahat\subSS$.

Also according to \eqref{eq:qcdot-Ak-b0},
elements in $\widehat{\bS}(\bBschat\subdelta)$ are asymptotically collinear.
For the vector
$$
\bv(c_1,\dots,c_K) = (c_1 \bBsc_0\trans, \dots, c_K \bBsc_0\trans)\trans,
$$
we have
\begin{align}
\bv(c_1,\dots,c_K)\trans \widehat{\bS}(\bBschat\subdelta)
= & \frac{2}{n}\sum_{i=1}^{n}\sum_{k=1}^K c_k \bBsc_0\trans\Abb_k \|\bbeta_0\|_2^{-1} \Bbb^{-1}\{\bZ_i-\bb(C_j)\}[\delta_i - g\{h_0(C_i) + \bbeta_0\trans\bZ_i\}] \notag \\
 & +\frac{2}{N}\sum_{i=1}^{N} \sum_{k=1}^K c_k \left\{\bBsc_0\trans\dot{\bq}_k(\bBsc_0)+ \bBsc_0\trans\dot{\bq}^c_k(C_i,\bBsc_0)\right\} + o_p(n^{-1/2} + N^{-1/2}) \notag \\
 & = o_p(n^{-1/2} + N^{-1/2}). \label{eq:Sbhat-b0}
\end{align}
By \eqref{eq:Sbhat-b0},
the rank of the asymptotic variance of $\sqrt{n}\widehat{\bS}(\bBschat\subdelta)$
is at most $K(p-1)$.

\subsection*{\underline{Consistency of $\Wbbhat(\Pbb)$}}
We identify the optimal projection as
\begin{align}
  \Wbb\subopt(\Pbb)\trans =  & \overline{\Cov}\left(\sqrt{n}\bbetahat\subdelta,\sqrt{n}\Pbb\widehat{\bS}(\bBschat\subdelta)\right)
  \overline{\Var}\left(\sqrt{n}\Pbb\widehat{\bS}(\bBschat\subdelta)\right)^{-1}\Pbb  \notag \\
  =  & \overline{\Cov}\left(\sqrt{n}\bbetahat\subdelta,\sqrt{n}\widehat{\bS}(\bBschat\subdelta)\right)
 \Pbb\trans \overline{\Var}\left(\sqrt{n}\Pbb\widehat{\bS}(\bBschat\subdelta)\right)^{-1}\Pbb  \notag \\
  = & \left\{
  \begin{array}{ll}
  \left\{\Bbb^{-1} \Sigbb\subdelta(\Bbb^{-1})\trans\Abb\trans\|\bbeta_0\|_2^{-1}
     +\frac{n}{N}\Sigbb\subdS\right\}\Pbb\trans & \\
     \quad\times
     \left[2\Pbb\left\{\Abb\Bbb^{-1} \Sigbb\subdelta(\Bbb^{-1})\trans\Abb\trans\|\bbeta_0\|_2^{-2}
  + \frac{n}{N}(\Sigbb\subS+\Abb\Sigbb\subdS+\Sigbb\subdS\trans\Abb\trans)\right\}
  \Pbb\trans \right]^{-1}\Pbb, & N \asymp n \\
  \Bbb^{-1} \Sigbb\subdelta(\Bbb^{-1})\trans\Abb\trans\Pbb\trans
  \left\{2\Pbb\Abb\Bbb^{-1} \Sigbb\subdelta(\Bbb^{-1})\trans\Abb\trans\Pbb\trans\right\}^{-1}\Pbb\|\bbeta_0\|_2,   & N \gg n
  \end{array}
  \right. , \label{def:Wopt}
\end{align}
where $\overline{\Cov}$ and $\overline{\Var}$ denote the asymptotic covariance and variance.
The term
$$
\Pbb\trans\overline{\Var}\left(\sqrt{n}\Pbb\widehat{\bS}(\bBschat\subdelta)\right)^{-1}\Pbb
= \Pbb\trans\left\{\Pbb\overline{\Var}\left(\sqrt{n}\widehat{\bS}(\bBschat\subdelta)\right) \Pbb\trans\right\}^{-1}\Pbb
$$
in \eqref{def:Wopt}
satisfies the definition of the generalized inverse for the rank deficient
$\overline{\Var}\left(\sqrt{n}\widehat{\bS}(\bBschat\subdelta)\right)$.

Since the loss \eqref{eq:lm_what} is convex, the solution is uniquely identified by the KKT condition,
\begin{equation}
  \widehat{\bw}_j(\Pbb) = \left\{ \frac{1}{B}\sum_{b=1}^B \Pbb\bShat\supb(\bBschat\subdelta\supb)\bShat\supb(\bBschat\subdelta\supb)\trans\Pbb\trans
  \right\}^{-1}\frac{1}{B}\sum_{b=1}^B \betahat\subdj\supb \Pbb\bShat\supb(\bBschat\subdelta\supb) .
\end{equation}
We have shown that $\sqrt{n}\betahat\subdj$ and $\sqrt{n}\bShat(\bBschat\subdelta)$ are asymptotically regular.
By \cite{JinYingWei01}, sample moments of perturbation samples converge to the asymptotic moments
at $n^{-1/2}$ rate.
We have
\begin{equation}
  \widehat{\bw}_j(\Pbb) =
  \overline{\Var}\left(\sqrt{n}\Pbb\widehat{\bS}(\bBschat\subdelta)\right)^{-1} \overline{\Cov}\left(\sqrt{n}\Pbb\widehat{\bS}(\bBschat\subdelta),\sqrt{n}\betahat\subdj\right)
  + O_p\left(n^{-1/2}\right).
\end{equation}
We assemble the estimated coefficients $\widehat{\bw}_1, \dots, \widehat{\bw}_{p}$
to construct estimated projection matrix
\begin{align}
  \Wbbhat(\Pbb) = & \Pbb\trans (\widehat{\bw}_1(\Pbb), \dots, \widehat{\bw}_{p}(\Pbb)) \notag \\
  = &\Pbb\overline{\Var}\left(\sqrt{n}\Pbb\widehat{\bS}(\bBschat\subdelta)\right)^{-1} \overline{\Cov}\left(\sqrt{n}\Pbb\widehat{\bS}(\bBschat\subdelta),\sqrt{n}\bbetahat\subdelta\right)
  + O_p\left(n^{-1/2}\right) \notag \\
  = &\Wbb\subopt(\Pbb) + O_p\left(n^{-1/2}\right).
\end{align}

\subsection*{\underline{Asymptotic distribution of $\bbetahat\subSS$}}

Now, we may give the asymptotic approximation of $\bbetahat\subSS$,
\begin{align}
 \bbetahat\subSS-\bbeta_0  = & \bbetahat\subdelta - \Wbbhat_{\sub comb}\trans \widehat{\bS}(\bBschat\subdelta) -\bbeta_0 \notag \\
 = &  \bbetahat\subdelta -\bbeta_0- \overline{\Wbb\subopt}\trans \widehat{\bS}(\bBschat\subdelta) + O_p\left(n^{-1}\right) \notag \\
 = & \frac{2}{n}\sum_{i=1}^{n}\left(\Ibb_p - \overline{\Wbb\subopt}\trans\Abb \|\bbeta_0\|_2^{-1}\right) \Bbb^{-1}\{\bZ_i-\bb(C_j)\}[\delta_i - g\{h_0(C_i) + \bbeta_0\trans\bZ_i\}] \notag \\
 & -\frac{2}{N}\sum_{i=1}^{N} \overline{\Wbb\subopt}\trans\left\{\dot{\bq}(\bBsc_0)+ \dot{\bq}^c(C_i,\bBsc_0)\right\} + o_p(n^{-1/2} + N^{-1/2}).
\end{align}
The asymptotic distribution of $\bbetahat\subSS$ is thus
\begin{equation}
  \sqrt{n}(\bbetahat\subSS - \bbeta_0) \leadsto
  N\left(\mathbf{0},4\Sigbb\subSS\right),  \tag{\ref{eq:SSL_norm}}
\end{equation}
with $\Sigbb\subSS$ defined next to \eqref{eq:SSL_norm} in the main text.

To obtained the simplified representation for $N\gg n$ scenario as in \eqref{eq:SSL-Nggn},
we shall show that
\begin{equation}\label{eq:SSL-Nggn-mat}
  \left(\Ibb_p - \Wbb\subopt(\Pbb)\trans\Abb \|\bbeta_0\|_2^{-1}\right)
   = \frac{\bbeta_0\bbeta_0\trans \Bbb\trans \Sigbb\subdelta^{-1}\Bbb}{\bbeta_0\trans \Bbb\trans \Sigbb\subdelta^{-1}\Bbb\bbeta_0}.
\end{equation}
Since $\overline{\Wbb\subopt}$ is a linear combination of $\Wbb\subopt(\Pbb_j)$ or $\Wbb\subopt(\Pbb_{k,j})$,
the identity \eqref{eq:SSL-Nggn-mat}
would imply
$$
  \left(\Ibb_p - \overline{\Wbb\subopt}\trans\Abb \|\bbeta_0\|_2^{-1}\right)
   = \frac{\bbeta_0\bbeta_0\trans \Bbb\trans \Sigbb\subdelta^{-1}\Bbb}{\bbeta_0\trans \Bbb\trans \Sigbb\subdelta^{-1}\Bbb\bbeta_0}.
$$
Recall that each $\Abb_k$
is a rank $p-1$ symmetric matrix, whose columns and rows
are orthogonal to $\bBsc_0$ according to \eqref{eq:qcdot-Ak-b0}.
The rows of $\Pbb \Abb$ are linear combinations of the rows of $\Abb_k$,
which should also be orthogonal to $\bBsc_0$.
By the design of $\Pbb$, the rank of $\Pbb\Abb$ equals $p-1$.
Hence, the rows of $\Pbb\Abb$ must form a basis of the linear subspace $\bbeta_0^\perp = \{\bv \in \Rbb^p: \bv\trans \bbeta_0 = 0\}$.
We may represent arbitrary $\bv_\perp \in \bbeta_0^\perp$ as
\begin{equation}
  \bv_\perp = \Abb\trans \Pbb\trans \bv^*_\perp, \;
  \bv^*_\perp = \left(\Pbb\Abb\Abb\trans \Pbb\trans\right)^{-1} \Pbb\Abb\bv_\perp.
\end{equation}
Recall that we have under $N\gg n$ case
$$
\Wbb\subopt\trans(\Pbb) = \overline{\Var}\left(\sqrt{n}\bbetahat\subdelta\right) \Abb\trans
 \Pbb\trans \left\{\Pbb\Abb\overline{\Var}\left(\sqrt{n}\bbetahat\subdelta\right)\Abb\trans\Pbb\trans\right\}^{-1}\Pbb
 \|\bbeta_0\|_2.
$$
We start with the left hand side,
\begin{align*}
 & \bv_\perp\trans \left(\Ibb_p - \Wbb\subopt(\Pbb)\trans\Abb \|\bbeta_0\|_2^{-1}\right) \\
 = &
\bv\strans_\perp\Pbb\Abb  - \bv\strans_\perp\Pbb\Abb\overline{\Var}\left(\sqrt{n}\bbetahat\subdelta\right) \Abb\trans
 \Pbb\trans \left\{\Pbb\Abb\overline{\Var}\left(\sqrt{n}\bbetahat\subdelta\right)\Abb\trans\Pbb\trans\right\}^{-1}\Pbb\Abb \\
=  & \mathbf{0}\trans.
\end{align*}
Thus, we have
$$
\bbeta_0^\perp  \subseteq \ker\left(\left(\Ibb_p - \Wbb\subopt(\Pbb)\trans\Abb \|\bbeta_0\|_2^{-1}\right)\trans\right).
$$
By the fundamental theorem of linear algebra,
the rank of $\left(\Ibb_p - \Wbb\subopt(\Pbb)\trans\Abb \|\bbeta_0\|_2^{-1}\right)$ is at most 1,
and it must admit the following one-dimensional singular value decomposition
\begin{equation}
  \Ibb_p - \Wbb\subopt(\Pbb)\trans\Abb \|\bbeta_0\|_2^{-1} = \bbeta_0 \sigma \br\trans
\end{equation}
for some $\sigma \in \Rbb$ and $\br \in \Rbb^p$.
To calculate $\sigma$ and $\br$, we consider
\begin{align*}
& \bbeta_0 \trans\overline{\Var}\left(\sqrt{n}\bbetahat\subdelta\right)^{-1}\bbeta_0 \sigma \br\trans \\
 = &
 \bbeta_0 \trans\overline{\Var}\left(\sqrt{n}\bbetahat\subdelta\right)^{-1} \left(\Ibb_p - \Wbb\subopt(\Pbb)\trans\Abb \|\bbeta_0\|_2^{-1}\right) \\
 = & \bbeta_0\trans \overline{\Var}\left(\sqrt{n}\bbetahat\subdelta\right)^{-1}
 - \bbeta_0\trans \overline{\Var}\left(\sqrt{n}\bbetahat\subdelta\right)^{-1}\overline{\Var}\left(\sqrt{n}\bbetahat\subdelta\right) \Abb\trans
 \Pbb\trans \left\{\Pbb\Abb\overline{\Var}\left(\sqrt{n}\bbetahat\subdelta\right)\Abb\trans\Pbb\trans\right\}^{-1}\Pbb\Abb \\
 = & \bbeta_0\trans \overline{\Var}\left(\sqrt{n}\bbetahat\subdelta\right)^{-1}.
\end{align*}
Then, we must have
$$
\sigma \br\trans = \frac{\bbeta_0\trans \overline{\Var}\left(\sqrt{n}\bbetahat\subdelta\right)^{-1}}{\bbeta_0 \trans\overline{\Var}\left(\sqrt{n}\bbetahat\subdelta\right)^{-1}\bbeta_0 }
= \frac{\bbeta_0\trans \Bbb\trans \Sigbb\subdelta^{-1}\Bbb}{\bbeta_0\trans \Bbb\trans \Sigbb\subdelta^{-1}\Bbb\bbeta_0}.
$$
Therefore, we have proved \eqref{eq:SSL-Nggn-mat}.

\end{document}